\documentclass[aps,pre,twocolumn]{revtex4-2}
\pdfoutput=1
\usepackage[utf8]{inputenc}
\usepackage{graphicx}
\graphicspath{{Figures for main/}}
\usepackage{float}
\usepackage{amssymb}
\usepackage{amsmath}
\usepackage{diagbox}
\usepackage{slashbox}
\usepackage{graphics}
\usepackage[utf8]{inputenc}
\usepackage{bm}
\usepackage{xcolor}

\definecolor{cornellred}{rgb}{0.7, 0.11, 0.11}
\definecolor{cadmiumred}{rgb}{0.89, 0.0, 0.13}

\newcommand{\gc}[1]{\textcolor{black}{#1}}

\newcommand{\xl}[1]{\textcolor{black}{#1}}

\begin{document}

\title{Feasibility and stability in large Lotka Volterra systems with interaction structure}

\author{Xiaoyuan Liu$^1$, George W.A. Constable, and Jonathan W. Pitchford}
\affiliation{Department of Mathematics, University of York}

\begin{abstract}

Complex system stability can be studied via linear stability analysis using Random Matrix Theory (RMT) or via feasibility (requiring positive equilibrium abundances). Both approaches highlight the importance of interaction structure. Here we show, analytically and numerically, how RMT and feasibility approaches can be complementary. In generalised Lotka-Volterra (GLV) models with random interaction matrices, feasibility increases when predator-prey interactions increase; increasing competition/mutualism has the opposite effect. These changes have crucial impact on the stability of the GLV model.

\end{abstract}


\maketitle

\section{Introduction}

In the 1950s, ecologists such as Odum and MacArthur argued~\cite{odum2013fundamentals,macarthur1955fluctuations} that ecosystems with a larger number of species tend to be more stable than less biodiverse systems. This idea was famously mathematised by May in 1972, who applied random matrix theory (RMT) to the problem~\cite{may1972will}. May considered perturbations in $n$ species abundances, $\bm{\zeta}$, linearised about a hypothetical fixed point, with near-equlibrium dynamics described by 
\begin{equation}
    \frac{ \mathrm{d} \bm{\zeta} }{ \mathrm{d} t } = A \bm{\zeta}
\label{maymodel}
\end{equation}
where he suggested parameterising $A$ according to 
\begin{equation}
   A_{ii}=-1 \,, \quad A_{ij} = \sigma c a_{ij}
\label{maymodelparam}
\end{equation}
with $A_{ii}$ representing the species self-regulation at equilibrium and $a_{ij} \sim \mathcal{N}(0,1)$ and $c \sim \mathrm{\bm{B}}(1,C)$. Here $A_{ij}$ represents random species interactions that are non-zero with probability $C$ (referred to as connectance) and when present have standard deviation $\sigma$ (referred to as interaction strength). Since the asymptotic stability of Eq.~(\ref{maymodel}) is governed solely by its eigenvalues, system-level stability is determined by characterising the eigenvalues of random matrix $A$.

The eigenvalue distribution of $A$ is uniform across a circle in the complex plane, centered on $(-1,0)$ and with radius $\sigma \sqrt{ n C}$ as $n\to\infty$~\cite{wigner_1958, may1972will, tao_vu_krishnapur_2010}.

Thus the stability criterion for Eq.~(\ref{maymodel}) is $\sigma\sqrt{nC}<1$ (see Fig.~1(a)). This suggests that more diverse ecosystems with more interspecific interactions are less likely to be stable for a given variance in interaction strength.

Allesina and Tang \cite{allesina_tang_2012} added ecologically-motivated structure to May's approach, choosing elements of $A$ pairwise by imposing a correlation, $\rho$, between $A_{ij}$ and $A_{ji}$ for $j\neq i$, 
\begin{eqnarray}
&&(A_{ij},A_{ji})=\sigma c(a_{ij},a_{ji})\hspace{2.5mm}\text{where}\label{A+Tparam}\\ &&(a_{ij},a_{ji}) \sim  \mathcal{N}(\bm{0},\Sigma)\hspace{2.5mm}\text{with}\hspace{2.5mm} \Sigma=\left[(1,\rho),(\rho,1) \right]\nonumber
\end{eqnarray}
where again $c\sim B(1,C)$. Ecologically, $\rho<0$ implies more predator-prey interactions in the ecosystem \gc{($A_{ij}$ and $A_{ji}$ are more likely to have opposite signs)}, while $\rho>0$ implies more mutualistic and competitive interactions \gc{($A_{ij}$ and $A_{ji}$ are more likely to have the same sign)}. Utilising another RMT result \cite{girko1986elliptic,sommers1988spectrum} they generalised May's stability criterion to 

\begin{equation}    \sigma\sqrt{nC}(1+\rho)<1 \,.
\label{girkomaywigner}\end{equation}

Thus, increasing the proportion of predator-prey interactions increases stability, whilst increasing the proportion of competitive and mutualistic interactions reduces stability in Eq.~\eqref{maymodel} (see Fig.~1(a)). Eq.~\eqref{girkomaywigner} implies that in the extreme limit $\rho\to -1$, ecosystems are stable as long as there is self-regulation.

These analytic results are independent of the underlying non-linear model \gc{from which they are hypothetically derived. However}, this apparent generality conceals an implicit assumption that the fixed point \gc{about which the non-linear system is linearised (to arrive at Eq.~\eqref{maymodel})} exists and is biologically meaningful. Such \gc{biologically meaningful} fixed points, where every species is present at a positive abundance, are termed feasible equilibria~\cite{roberts1974stability}.

We use the generalised Lotka-Volterra model (GLV)
\begin{equation}
    \frac{ \mathrm{d} \bm{x} }{ \mathrm{d} t} = \bm{x} \odot \left( \bm{r} + A \bm{x} \right) \,,
\label{GLV}
\end{equation}
\gc{to explore the links between the parameterisations of the interaction matrix $A$ in Eqs.~(\ref{maymodelparam}-\ref{A+Tparam})} and feasibility. Here $x_i$ is the abundance of species $i$, $r_i$ is its intrinsic growth rate, $A$ the interaction matrix, and $\odot$ the Hadamard product. Eq.~(\ref{GLV}) has a single non-zero fixed point, $\bm{x}^*$, with a Jacobian, $J$, such that
\begin{equation}
    \bm{x}^{*}=-A^{-1}\bm{r} \,, \qquad J=\mathrm{diag}(\bm{x}^{*})A
\label{xfp}
\end{equation}
\xl{Note that if the elements of $A$ are drawn from a random distribution, then $\bm{x}^{*}$ is also a random variable (see, for instance Fig.~\ref{x1_x2}). We denote the multivariate distribution of $\bm{x}^{*}$ as $P(\bm{x}^{*})$. In particular, there is nothing intrinsic about the structure of $\bm{x}^{*}$ in Eq.~\eqref{xfp} that guarantees that it is feasible (\gc{i.e. that $x^*_i>0\,\forall\,i$}). \gc{Instead, for any given randomly sampled $A$, there is a probability that the fixed point is feasible, which we denote $P_{feas}$}.} The relationships between feasibility, stability and different system constraints such as interaction structure is a central theme in theoretical ecology~\cite{bunin2017ecological}.

Early analytic insight into the feasibility of $\bm{x}^*$ in Eq.~\eqref{xfp} assumed that $A$ had interaction coefficients with fixed strengths, or with randomly generated signs~\cite{roberts1974stability,gilpin1975stability,goh1977feasibility}. Stone \cite{stone1988some} linked this to May's approach by considering the probability that $\bm{x}^*$ is feasible given an ensemble of random interaction matrices parameterised \xl{according to} Eq.~(\ref{maymodelparam}).  Under the condition that $r_i=1$ $\forall$ $i \in[i,n]$, Stone assumed that such a parameterisation of interaction matrices gives rise to a normally distributed $x^{*}_{i}$ (see Figure \ref{x1_x2} \xl{and Supplemental Material Section VIII}).

\begin{figure}[t!]
    \centering
    \includegraphics[width=0.49\textwidth]{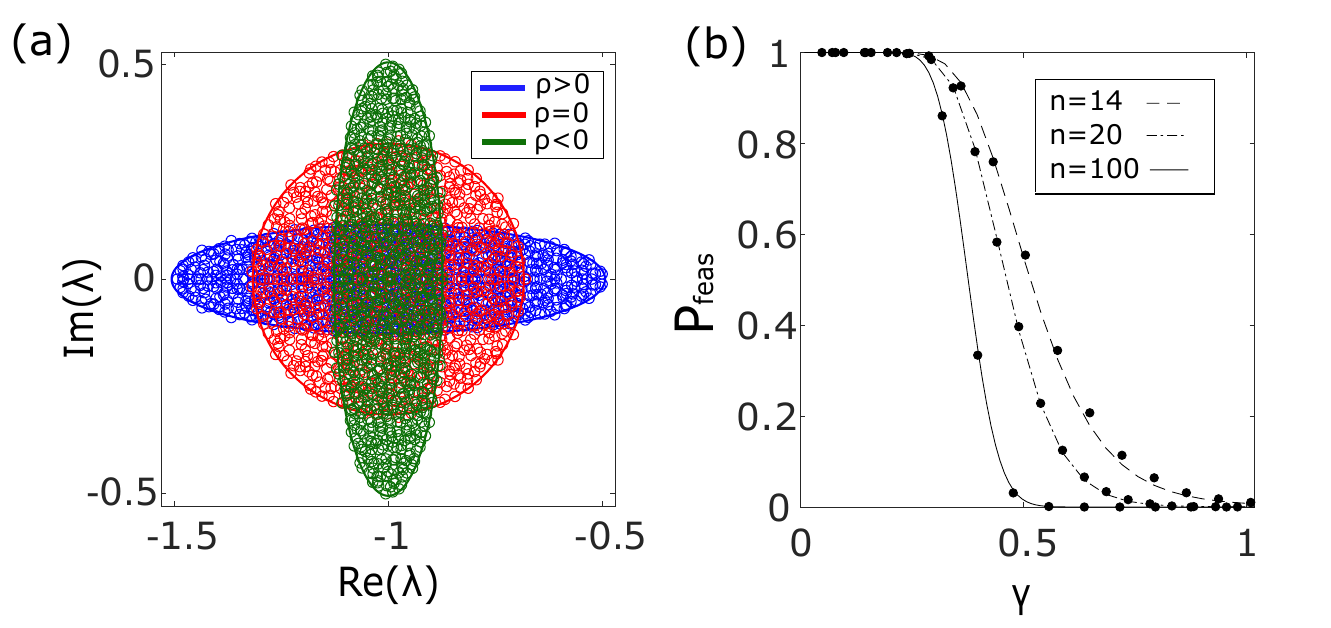}
    \caption{Panel~(a): Eigenvalue distributions of interaction matrix $A$ \gc{parameterised according to Eq.~\eqref{maymodelparam} (red, $\rho=0$, see~\cite{may1972will}) and Eq.~\eqref{A+Tparam} (blue and green, $\rho\neq 0$, see~\cite{allesina_tang_2012})}, used to infer the stability of the linear model proposed in Eq.~\eqref{maymodel}.
Parameter values are $\sigma=0.01$, $n=1000$, $C=1$ and $|\rho|=0.6$. Panel~(b): Feasibility probability, $P_{\mathrm{feas}}$, \gc{for an ensemble of random fixed points from the non-linear GLV model, Eq.~\eqref{GLV}, with interaction matrices parameterised according to Eq.~\eqref{maymodelparam} ($\rho=0$, see~\cite{stone1988some}). $P_{\mathrm{feas}}$ is plotted as a function of May's complexity parameter $\gamma=\sigma \sqrt{nC}$, for community sizes ranging from $n=14$ to $n=100$}. \xl{In this panel $C=1$.} Curves are analytical predictions and markers are \xl{numerical simulations, obtained by sampling $10^4$ random interaction matrices $A$ parameterised according to Eq.~\eqref{maymodelparam} and calculating the proportion of those that give rise to a feasible equilibrium solution of the GLV model (see Supplemental Material IV)}.}
    \label{May_Allesina} 
\end{figure} 

Stone showed that for a fully connected system $C=1$, the probability of feasibility is
 \begin{equation}
    P_{\mathrm{feas}}=2^{-n}\bigg(1+\text{erf}(\frac{1}{\gamma_S\sqrt{1+\gamma_S^2+\gamma_S^4})})\bigg)^n \,,
\label{pfeasanalytialfy}
\end{equation}
where $\gamma_S=\sigma\sqrt{n}$ is known as the disturbance in Stone's analysis, \xl{which is equivalent to May's definition of complexity for the case $C=1$}. We see that $P_{\mathrm{feas}}$ drops sharply at a critical value of $\gamma_S$, and also has an additional dependence on system size $n$ (see Fig.~1(b)). By working in the limit $n\to\infty$, \cite{clenet2022equilibrium, bizeul2021positive} determined a threshold interaction strength above which feasibility is lost in GLV models with \xl{interaction matrices parameterised according to Eq.~\eqref{maymodelparam}}. An analytical prediction for the relationship between $P_{\mathrm{feas}}$ and the complexity $\gamma=\sigma\sqrt{nC}$ which accounts for $C$ was obtained by Dougoud et al. \cite{dougoud2018feasibility}. Akjouj et al. \cite{akjouj2021feasibility} investigated the feasibility of sparse ecosystems with interaction matrices that are block structured and d-regular (where each species interacts with d other species). Together these results suggest that feasibility is the more critical measure of complex system stability; compared to linear stability, feasibility is lost at smaller values of complexity.

Here we seek to strengthen the links between RMT~\cite{may1972will,allesina2015stability} and feasibility analyses by calculating how the feasibility of an ecosystem changes with complexity~\cite{stone1988some,stone2018feasibility,dougoud2018feasibility,akjouj2021feasibility} when additional species interaction structure is \xl{accounted for}~\cite{allesina_tang_2012,allesina2015stability}. It was shown by Bunin \cite{bunin2017ecological} that feasible systems lose stability above a certain interaction strength by transition to a phase with multiple attractors. The interaction strength of this phase transition increases as predator-prey interactions increase. Numerical results by Clenet et al. \cite{clenet2022equilibrium} also show that systems biased towards predator-prey interactions lose feasibility at larger interaction strengths than \xl{systems without interaction structure}, and those biased towards competition and mutualism lose feasibility at smaller interaction strengths \xl{than systems without interaction structure}. They also obtained an analytical result for the interaction strength above which feasibility is lost, in the limit of large $n$. \gc{In this limit the effect of the correlation parameter $\rho$,} \xl{the parameter that governs the proportion of predator-prey or competition/mutualistic interactions}, disappears~\cite{clenet2022equilibrium}. In this paper, we instead work in the large but finite $n$ limit in order to explore the effect of $\rho$ on \xl{the probability of feasibility, $P_{\mathrm{feas}}$. In order to calculate $P_{\mathrm{feas}}$, we must also obtain an approximation for the distribution of fixed points.} This approximation opens up the possibility of leveraging recent results~\cite{gibbs2018effect,baron2022eigenvalues} to determine the probability of stability of the GLV model with interaction structure.

\begin{figure*}[ht!]
    \centering
    \includegraphics[width=160mm]{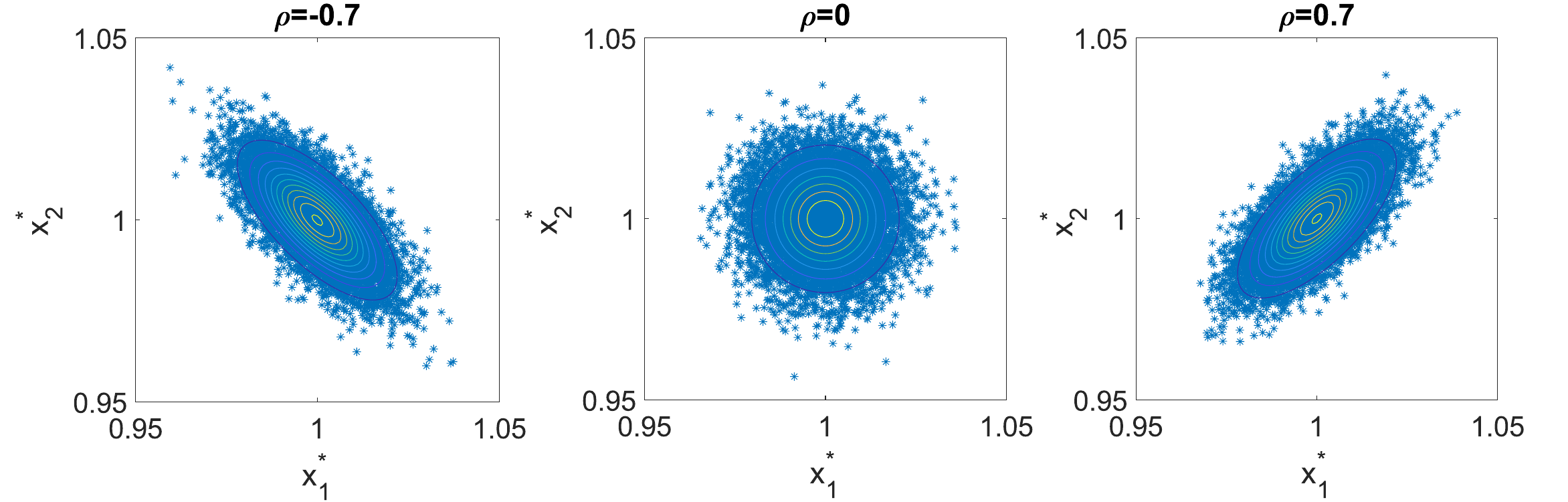}
    \caption{Plots showing the joint distribution of $x^{*}_{1}$ and $x^{*}_{2}$ for the GLV model Eq.~\eqref{GLV} with $n=2$, $\sigma=0.01$ and $C=1$. Blue markers represent $10^4$ numerical solutions of the GLV model, obtained as described in Supplemental Material IV. Contours are analytical predictions for the joint distribution of $x^{*}_{1}$ and $x^{*}_2$ calculated using Eqs.~(\ref{exx}-\ref{cxxx}).}
    \label{x1_x2}
\end{figure*}

\section{Analysis} 
Following Stone~\cite{stone2016google} we obtain an analytical approximation of $P_{\mathrm{feas}}(\gamma)$ via the distribution of equilibrium species abundances $P(\bm{x}^{*})$. \xl{In particular Stone \cite{stone1988some} applied the Central limit theorem to $\bm{x}^{*}$ in Eq. \eqref{neuman} to argue that $P(\bm{x}^{*})$ is normal as $n\to\infty$, and this normality remains a good approximation when $n$ is large but finite (see Supplemental Material Section VIII).} The task of calculating the feasibility probability is then equivalent to calculating 

\begin{equation}
    P_{\mathrm{feas}}= \int_{\bm{x}^*=\bm{0}}^{\infty} P(\bm{x}^{*}) \mathrm{d} \bm{x}^* \approx \int_{\bm{x}^*=\bm{0}}^{\infty} \mathcal{N}( \bm{\mu}_{\bm{x}^*} \Sigma_{\bm{x}^*}) \mathrm{d} \bm{x}^* \,
\label{pfeasanalytialfya}
\end{equation}
where $\bm{\mu}_{\bm{x}^*}$ and $\Sigma_{\bm{x}^*}$ are respectively the mean and covariance matrix of the species abundances at equilibrium. \gc{Note that by symmetry, we can see that for interaction matrices randomly generated according to Eq.~\eqref{A+Tparam}, $\bm{\mu}_{\bm{x}^*}$ and $\Sigma_{\bm{x}^*}$ are themselves highly symmetric, with $[\bm{\mu}_{\bm{x}^*}]_i=[\bm{\mu}_{\bm{x}^*}]_j$, $[\Sigma_{\bm{x}^*}]_{ii}=[\Sigma_{\bm{x}^*}]_{jj}$ and $[\Sigma_{\bm{x}^*}]_{ij}=[\Sigma_{\bm{x}^*}]_{ji}$ for all $i,j\,\in [1,n]$ (i.e. $\bm{\mu_{x}^{*}}$ is a constant vector and the variance-covariance matrix $\Sigma_{\bm{x}^{*}}$ is a double constant matrix \cite{o2021double}).}

We now calculate approximations for $\bm{\mu}_{\bm{x}^*}$ and $\Sigma_{\bm{x}^*}$. For simplicity we focus on the case $r_i=1$ $\forall$ $i$ in Eq.~(\ref{GLV}). Recall that following~\cite{allesina2015stability}, the elements of the interaction matrix $A_{ij}$ and $A_{ji}$ have correlation $\rho$. Writing $A=\sigma\mathcal{E}-\mathbf{I}$, our fixed point in Eq.~(\ref{xfp}) can be expressed as a Neumann series~\cite{kress_2014} for $||\sigma\mathcal{E}||<1$:

\begin{equation}
    \bm{x}^{*}=(\mathbf{I}-\sigma\mathcal{E})^{-1}\mathbf{r} \equiv \bigg(\sum^{\infty}_{j=0}(\sigma\mathcal{E})^{j}\bigg)\mathbf{r} .
\label{neuman}
\end{equation}
This enables us, in principle, to calculate $x^{*}_{i}$ up to an arbitrary order in $\sigma$. In our work, we approximate $E(x^{*}_{i})$, $Var(x^{*}_{i})$ and $Cov(x^{*}_{i},x^{*}_{j})$ taking into account $\rho$ and $C$. Using Eq. \eqref{neuman}, we approximate $E(x^{*}_{i})$ and $Var(x^{*}_{i})$ up to and including order $\sigma^6$. \xl{Using the fact that the product of an odd number of normal random variables with zero mean have zero expectation, we know that all terms of $E(x^{*}_{i})$ at odd orders of $\sigma$ vanish. From Eq. \eqref{neuman}, we find that the expression for $x^{*}_{i}$ at this given order is}

\begin{flalign}
    E(x^{*}_{i})=E\left(1+\sigma^2\sum^{n}_{\substack{j=1\\j\neq i}}\sum^{n}_{\substack{k=1\\k\neq j}}\kappa a_{ij}a_{jk}\right)+e_4\sigma^4+e_6\sigma^6\label{EXepression}\end{flalign} \xl{where $e_4$ and $e_6$ are coefficients of $\sigma^4$ and $\sigma^6$ respectively in the expectation of $x^{*}_{i}$, and \begin{equation}
    \kappa=\begin{cases}
C &  \text{if}\quad i=k \,, \\
C^2 & \text{if}\quad i\neq k\,,
\end{cases}
\end{equation}
since $i=k$ corresponds to the case where $a_{jk}=a_{ji}$, which corresponds to the case where $A_{ij}$ and $A_{ji}$ are both nonzero with probability $C$ (see Eq.~\eqref{A+Tparam} and Allesina and Tang \cite{allesina2015stability}). We use Eq.~\eqref{EXepression} to illustrate how we obtain our approximation of $E(x^{*}_{i})$. Since $E(a_{ij}a_{ji})=\rho$, $E(a_{ij})=0$ and $E(a_{ij}a_{jk})=0$ if $k\neq i$, Eq.~\eqref{EXepression} is equal to}
\begin{equation}
    E(x^{*}_{i})=1+(n-1)\rho C\sigma^2+e_4\sigma^4+e_6\sigma^6\label{exx}
\end{equation} \textcolor{black}{where through direct calculation, it can be shown that $e_4=(n-1)(C+\rho^2(2C+2C^2(n-2)))$, given by Eq.~(S12). Similarly we can calculate $e_6$, which is given by Eq.~(S53) of the Supplemental Material.}

\gc{An analogous approach can be used to obtain an approximation for $Var(x^{*}_{i})$ and $Cov(x^{*}_{i},x^{*}_{j})$ (see Supplemental Material Section I),} \xl{with $Var(x^{*}_{i})$ given by}
\begin{eqnarray}
    Var(x^{*}_{i})=(n-1)C\sigma^2+v_4\sigma^4+v_6\sigma^6+O(\sigma^8)
\label{vxxx}
\end{eqnarray} 
where $v_4$ and $v_6$ are the coefficients of $\sigma^4$ and $\sigma^6$ respectively, which depend on $n$, $\rho$ and $C$. Specifically, $v_4$ is the coefficient of $\sigma^4$ in Eq.~(S20) 
and $v_6$ is given by Eq.~(S60) in the Supplemental Material. The formulas for $v_4$ and $v_6$ are too lengthy to produce here, however of particular note is the fact that they\gc{, along with coefficients $e_4$ and $e_6$,} are nontrivial polynomials that do not preserve the simple dependence on \xl{the complexity parameter $\gamma$ \gc{observed in~\cite{may1972will} or~\cite{allesina_tang_2012}}.} $Cov(x^{*}_{i},x^{*}_{j})$ is given by
\begin{equation}
    Cov(x^{*}_{i},x^{*}_{j})=\rho C\sigma^2+c_4\sigma^4+O(\sigma^6)
\label{cxxx}
\end{equation}
\xl{where $c_4=(3+(6+C(5n-11))\rho^2)$. \gc{While we could extend this approximation to order $\sigma^6$, we note that this makes little quantitative difference to the approximation.} In the expression for $Cov(x^{*}_{i},x^{*}_{j})$, the coefficient of each order of $\sigma$ is a factor of $n$ smaller than the corresponding coefficients in the expression for $E(x^{*}_{i})$ and $Var(x^{*}_{i})$ (see Supplemental Material Section VII). This implies that for a fixed value of large but finite $n$, $Cov(x^{*}_{i},x^{*}_{j})$ increases more slowly with $\sigma$ than $E(x^{*}_{i})$ and $Var(x^{*}_{i})$, and thus $Cov(x^{*}_{i},x^{*}_{j})$ plays a smaller role in governing how \gc{$P(\bm{x}^*)$, and similarly} $P_{\text{feas}}$, varies with $\sigma$. It is therefore possible to approximate $Cov(x^{*}_{i},x^{*}_{j})$ to order $\sigma^4$ without sacrificing the accuracy of the analytical prediction of $P_{\text{feas}}$. The slower increase in $Cov(x^{*}_{i},x^{*}_{j})$ with $\sigma$ is verified numerically in Figure S7. Since an analytical approximation of $Cov(x^{*}_{i},x^{*}_{j})$ to order $\sigma^6$ requires considerably more algebra (see Supplemental Material Section ID5) without conferring significant improvements to the accuracy of $P_{\text{feas}}$, we restrict our analysis to the order $\sigma^4$ approximation given in Eq.~\eqref{cxxx}. }

Eqs. (\ref{exx}-\ref{cxxx}) are then used to construct $\bm{\mu}_{\bm{x}^*}$ and $\Sigma_{\bm{x}^*}$ in Eq. \eqref{pfeasanalytialfya}. \gc{Note that we expect our approximation to hold when $n$ is large (such that $P(\bm{x}^*)$ is approximately normal, see Eq.~\eqref{pfeasanalytialfya}) and when $\sigma$ is small (such that the expansions in Eqs.~(\ref{exx}-\ref{cxxx}) remain sufficient). When these conditions are not met, the approximations given in Eqs.~(\ref{exx}-\ref{cxxx}) break down at lower values of $|\rho|$. For instance in a 25 species ($n=25$) system, the analytical approximation of $Var(x^{*}_{i})$ in Eq.~\eqref{vxxx} loses accuracy when $|\rho|>0.25$), while for a 100 species system $Var(x^{*}_{i})$ remains accurate up to $|\rho|=0.5$ (see Supplemental Material II)}.

\begin{figure}[t!]
    \centering
    \includegraphics[width=0.434\textwidth]{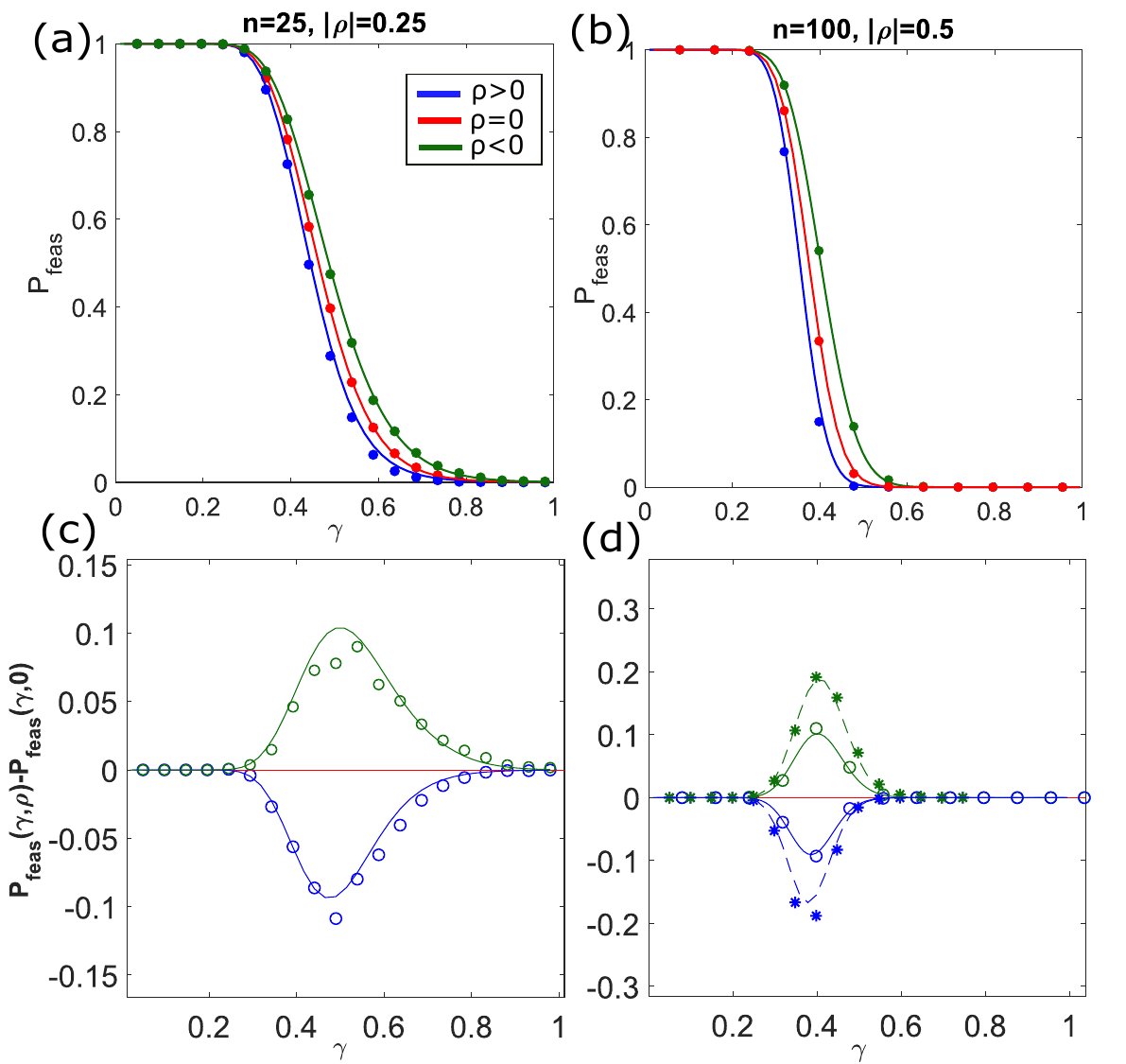}
        \caption{Panels (a) and (b) plot the feasibility probability $P_{\mathrm{feas}}$ as a function of complexity $\gamma$ for systems with ecologically motivated interaction structure\gc{: blue ($\rho>0$) biased toward competitive/mutualistic interactions; red ($\rho=0$) unbiased interactions; green ($\rho<0$) biased towards predator-prey interactions}. \gc{Panels }~(c)~and~(d) plot  the difference between $P_{\mathrm{feas}}$ in systems with $\rho\neq 0$ and $P_{\mathrm{feas}}$ in systems where $\rho=0$ ($P_{\mathrm{feas}}(\gamma,\rho)-P_{\mathrm{feas}}(\gamma,0)$) as a function of $\gamma$\gc{, with lines the prediction derived from Eq~\eqref{fhpercube} and markers the results of numerical simulation. In panel~(c), $n=25$ and hollow circles show the results of numerical simulations for the case $|\rho|=0.25$. In panel~(d), where $n=100$ (and our approximations are valid for larger values of $\rho$) hollow circles again represent the case the case $|\rho|=0.25$, while asterisks are numerical simulations for the case  $|\rho|=0.5$.} \xl{Numerical simulations are obtained by sampling $10^4$ random interaction matrices $A$ parameterised according to Eq. \eqref{maymodelparam} and calculating the proportion of those that give rise to a feasible equilibrium solution of the GLV model Eq. \eqref{GLV} (see Supplemental Material IV)}
        }
    \label{Pfeas_gamma_rho}
\end{figure}

\xl{The fact that our normal distributions feature such a high degree of symmetry, with $\bm{\mu}_{x^{*}}$ a constant vector and $\Sigma_{\bm{x}^{*}}$ a double constant matrix, allows us to further simplify the calculation of $P_{\mathrm{feas}}$. This provides ease of computation for large systems. Using the results of \cite{curnow1962numerical} which expresses integrals over the cubic region of the variable space, Eq. \eqref{pfeasanalytialfya} can be reduced to an expression involving a single integral}, given by \begin{equation}
    P_{\text{feas}}=\int^{\infty}_{-\infty}\bigg\{\prod^{n}_{i=1}\Phi(\frac{y_i-b_iu}{(1-b_i^2)^{1/2}})\bigg\}\phi(u)du
\label{fhpercube}\end{equation}
where $\phi(u)$ is the density function of a standard normal random variable $u$ and $\Phi(v)$ denotes the cumulative distribution function of a standard normal random variable $v$. In our analytical prediction of $P_{\text{feas}}$, we have that $y_i=\frac{E(x^{*}_i)}{\sqrt{Var(x^{*}_{i})}}$ and $b_i=\frac{\sqrt{Cov(x^{*}_{i},x^{*}_{j})}}{Var(x^{*}_{i})}$ \xl{(see Supplemental Material III)}. In other words, $P_{\text{feas}}$ is the expression obtained by substituting these expressions for $y_i$ and
$b_i$ into \eqref{fhpercube}. (see Supplemental Material III). Interestingly, note that in the results of~\cite{may1972will, allesina2015stability}, $C$ appears as a compound parameter with $\sigma^2$, but in Eqs.~(\ref{exx}-\ref{cxxx}), $C$ appears in a complicated polynomial form. The analytical prediction of $P_{\mathrm{feas}}(\gamma)$ is shown in Figure \ref{Pfeas_gamma_rho} (a)-(b). \xl{Moreover, the fact that $Cov(x^{*}_{i},x^{*}_{j})$ is a factor of $n$ smaller than $Var(x^{*}_{i})$ partly explains the observation of Clenet \cite{clenet2022equilibrium} that as $n\to\infty$, the effect of $\rho$ on $P_{\text{feas}}$ completely disappears. }

\section{Results}

\subsection{Predator-prey interactions increase the feasibility of random ecosystems}

The qualitative difference in how $P_{\mathrm{feas}}$ changes with the complexity $\gamma$ as the correlation $\rho$ is varied is shown analytically in Figure \ref{Pfeas_gamma_rho}. For a given value of $n$, when $\rho$ is positive (blue), feasibility is lost at a smaller complexity compared to the case where $\rho=0$ (red). However when $\rho$ is negative (green), we observe the opposite effect whereby feasibility is lost at a larger complexity than the case $\rho=0$.

It can be seen in Figure \ref{Pfeas_gamma_rho} that the magnitude of the difference between $P_{\mathrm{feas}}(\gamma,\rho)$ and $P_{\mathrm{feas}}(\gamma,0)$ also varies with $\gamma$. For instance when $\gamma$ is sufficiently small, there is no difference between $P_{\mathrm{feas}}(\gamma,\rho)$ and $P_{\mathrm{feas}}(\gamma,0)$, since $P_{\mathrm{feas}}$ is 1 regardless of $\rho$. The bottom panels of Figure \ref{Pfeas_gamma_rho} below plot this difference, demonstrating how it varies with $\gamma$. The difference between $P_{\mathrm{feas}}(\gamma,\rho)$ and $P_{\mathrm{feas}}(\gamma,0)$ is the greatest for intermediate values of complexity $\gamma$, where the system is transitioning rapidly away from feasibility. For a given system size $n$, the magnitude of this difference ($|P_{\mathrm{feas}}(\gamma,\rho)$-$P_{\mathrm{feas}}(\gamma,0)|$) also increases with the magnitude of $\rho$.

In Supplemental Material I.E, we see that for all values of $\rho$, the loss of feasibility in the GLV model with Allesina and Tang type interaction matrices occurs at a smaller complexity than the loss of stability in the corresponding linear model. As an extreme example, in linear systems comprising all predator-prey interactions ($\rho=-1$) stability is guaranteed regardless of ecosystem complexity (see Eq. \eqref{girkomaywigner}); conversely feasibility is still lost above a critical value of the complexity parameter $\gamma$ (see Figure S2 of Supplemental Material). Figure \ref{Pfeas_gamma_rho} demonstrates that the analytical results in Eq.~(\ref{exx}-\ref{cxxx}) can be used to accurately predict $P_{\text{feas}}$ as a function of $\gamma$ in the case where $C=1$. Furthermore, Supplemental Material V shows that the same analytical results remain highly accurate for predicting $P_{\text{feas}}$ as a function of $\gamma$ in the case where $C=0.3$. By comparing the feasibility probabilities of such a system with that of a fully connected system, we see that a sparsely connected system of $n=100$ shows an almost identical feasibility-complexity relation as a fully connected system. 

\xl{Most importantly, in Eqs.~(\ref{exx}-\ref{cxxx}) we have analytically approximated the distributions of $x^{*}_i$ for non-linear GLV models Eq.~\eqref{GLV} where the underlying interaction matrix $A$ is constructed according to Eq.~\eqref{A+Tparam}. This opens up the possibility to extend these results to predict the stability of GLV models with ecologically motivated interaction structures. Such a stability analysis is beyond the scope of this work, but would be attainable through detailed analysis of the GLV Jacobian. In the next section we investigate how this might be achieved within the scope of existing methods.} 

\subsection{Comparing RMT predictions with GLV Jacobian matrices}
Gibbs et al. \cite{gibbs2018effect} studied the eigenvalue distribution of a matrix that is assumed to be of the same structure as the GLV Jacobian (Eq.~\eqref{xfp} right), where $J$ is decomposed into a product of an interaction matrix $A$ and fixed points $\bm{x}^{*}$. However, for simplicity, they assume that the distribution from which $\bm{x}^{*}$ is drawn is independent of $A$, whereas this is clearly not the case (\xl{see} Eq.~\eqref{xfp} left).

\begin{figure*}[ht!]
     \centering
     \includegraphics[width=135mm]{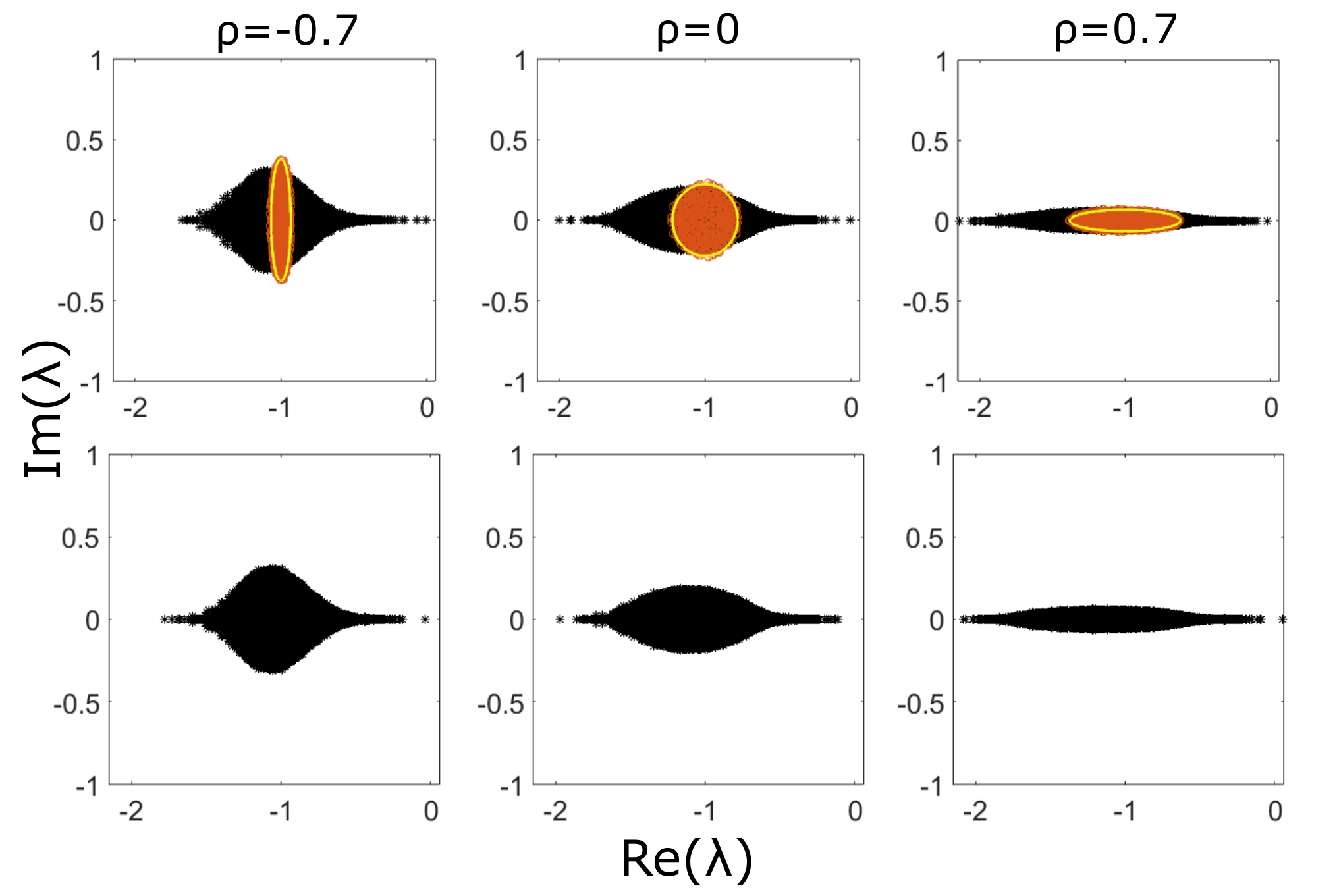}
     \caption{Top row: Orange ellipses are eigenvalue distributions of $A$ \xl{where $A$ is parameterised according to Eqs. (\ref{maymodelparam}-\ref{A+Tparam})}. Yellow boundaries are predicted by Allesina and Tang. Black markers represent 50 realisations of the eigenvalue distribution of the GLV Jacobian $J=\bm{x}^{*}A$ where the exact $\bm{x}^{*}$ corresponding to each given $A$ is used. Bottom row: 50 realisations of the eigenvalue distribution of $J=\bm{x}^{*}A$ where elements of $\bm{x}^{*}$ are sampled independently of $A$, from the multivariate normal \xl{distribution characterised by Eqs. (\ref{exx}-\ref{cxxx})}. Parameter values are $\sigma=0.01$, $n=500$ and $C=1$. Given these parameters, Eqs.~(\ref{exx}-\ref{cxxx}) predict that in the left panel $P_{\mathrm{feas}}=0.993$, middle panel $P_{\mathrm{feas}}=0.997$ and right panel $P_{\mathrm{feas}}=1.000$.}
     \label{verifying_Grilli}
 \end{figure*}

Gibbs' assumption of independence between the random elements of $A$ and $\bm{x}^{*}$ means that cross correlations between them need-not be considered, thereby simplifying the analysis. We test whether this assumption holds, in order to determine whether Gibbs' method may be applicable to calculating the eigenvalue distribution of the GLV \xl{Jacobian (Eq.~\eqref{xfp})}. To do so, we \xl{first} calculate the eigenvalue distribution of $J=\bm{x}^{*}A$ where the elements of $\bm{x}^{*}$ are sampled independently to those of $A$. \xl{The distribution from which we sample the elements of $\bm{x}^{*}$ is a normal distribution with $E(x^{*}_i)$, $Var(x^{*}_i)$ and $Cov(x^{*}_i,x^{*}_j)$ given by Eq.~(\ref{exx}-\ref{cxxx}), which we approximated. \textcolor{black}{$A$ is constructed according to Eq.~\eqref{A+Tparam}}}. We then compare this eigenvalue distribution \xl{(shown in Figure \ref{verifying_Grilli} bottom panels)} to that of the GLV Jacobian where the exact $\bm{x}^{*}$ corresponding to each given $A$ is used \xl{(shown in black markers of Figure \ref{verifying_Grilli} top panels)}.
 
\xl{By comparing the black markers on the top panels with those of the bottom panels of Figure \ref{verifying_Grilli}, we see that our method of sampling $\bm{x}^{*}$ independently of $A$ from our distribution of $\bm{x}^{*}$ works well in predicting the eigenvalue distribution of the GLV Jacobian. This comparison is conducted in a region where feasibility is almost surely guaranteed.} From the top panels, we see that when the correlation parameter is negative i.e $\rho<0$, the bulk eigenvalue distribution of $J$ gets stretched in the $Im(\lambda)$ plane, and when $\rho>0$ in the $Re(\lambda)$ plane. This qualitative effect is consistent with the result of Allesina and Tang \cite{allesina2015stability}. It is shown numerically in Supplemental Material VI that increasing $\rho$ decreases the average resilience of the GLV model.

The average maximum outlier eigenvalue \gc{(averaged over multiple realisations of the interaction matrix $A$)} is also correctly predicted by our theory, which relies on the assumption of statistical independence between $A$ and our calculated distribution of $\bm{x}^{*}$ (see Eqs. (\ref{exx}-\ref{cxxx})), as illustrated in Figure S6 (a). \xl{However, our theory does not correctly predict the maximum outlier eigenvalue of individual realisations of the GLV Jacobian. This suggests that cross-correlations between the entries of $A$ and $\bm{x}^{*}$ may be quantitatively important in calculating the stability of individual realisations of the GLV model.} As the stability of a system is governed solely by the eigenvalue with the largest real part, a stability analysis of the GLV model must be preceded via calculating such an eigenvalue. Below, we provide an insight into some possible techniques for calculating the stability of the GLV model \xl{with Allesina and Tang type interaction matrices}.

Stone \cite{stone2018feasibility} showed that provided that $||\sigma\mathcal{E}||$ is sufficiently small, the eigenvalue with the largest real part (outlier eigenvalue of $J$) is approximately equal to minus the abundance of the least abundant species i.e $\lambda_{max}\approx -\text{min}_{i\in\{1,n\}}x^{*}_{i}$; in which case we have the weak condition whereby feasibility corresponds to the local asymptotic stability of the GLV model. In the case where $\rho=0$ or $|\rho|$ is small, $-\text{min}_{i\in\{1,n\}}x^{*}_{i}$ is an accurate estimate of the outlier eigenvalue of $J$, however this accuracy breaks down as we increase $|\rho|$ (see Supplemental Material VI). 

Relying on Gibbs' assumption allows us to accurately capture the bulk eigenvalue distribution of $J$ and the effect that the correlation parameter $\rho$ has on the average \xl{resilience} over a large number of realisations \xl{(see Figure S6 (a))}, although it fails to accurately calculate the outlier eigenvalue of $J$ corresponding to a specific realisation of $A$.

\section{Discussion}
We have obtained an analytical prediction of the feasibility probability as a function of complexity $\gamma=\sigma\sqrt{nC}$ for random GLV models with interaction matrices of Allesina and Tang type \cite{allesina2015stability}. By extending the analytical result of \cite{clenet2022equilibrium} to the case of \xl{large, but} finite $n$, we have shown that a positive value of $\rho$ reduces the feasibility probability for a given complexity, while a negative value of $\rho$ increases the corresponding feasibility probability, an effect not quantifiable in the infinite $n$ limit. We have also accounted for the connectance $C$. Since natural ecological systems are sparsely connected \cite{gardner1970connectance}, both these generalisations mentioned above add biological realism to the result of Stone 2016 \cite{stone2016google}. Relationships between complexity and feasibility have also been studied by \cite{grilli2017feasibility}, where they characterised feasibility by how freely one could choose the intrinsic growth rate vectors to allow the system to remain feasible. As a whole, these results strengthen connections between feasibility and RMT systems, whilst also adding biological realism.

Along the way, we managed to analytically approximate the distribution of $\bm{x}^{*}$ as a function of the system parameters $n$, $C$, $\sigma$ and $\rho$. \xl{In doing so, we emphasise how the small covariance between the abundances of species can partly explain the observation of \cite{clenet2022equilibrium} that the effect of interaction structure on feasibility completely disappears as $n\to\infty$. Most importantly, our approximation of the distribution of $\bm{x}^{*}$} has allowed us to check the utility of Gibbs' assumption of independence between $\bm{x}^{*}$ and $A$ in predicting the eigenvalue distribution of the \xl{GLV Jacobian for systems with Allesina and Tang type interaction matrices}~\cite{stone2018feasibility, gibbs2018effect}. Figure \ref{verifying_Grilli} shows that Gibbs' assumption can be \xl{used} to accurately predict the effect of interaction structure \cite{allesina2015stability} on the eigenvalue distribution of feasible random GLV models. However, relying on this assumption does not allow us to \xl{accurately} calculate the outlier eigenvalue of the GLV Jacobian \xl{for} a particular realisation.

It is of note that our method for calculating the feasibility probability relies on several assumptions on the parameter values to ensure accuracy (see Supplemental Material I.E and II). We also assumed that $x^{*}_{i}$ is normally distributed. Since the Neumann series approximation for $x^{*}_{i}$ is normal in the limit $n\to\infty$, and is convergent if and only if $\sigma\sqrt{nC}<1$, our method is accurate for large $n$ and small $\sigma$ (see Supplemental Material VIII). Since the Neumann series expansion is precise, it is straightforward to extend our analysis to arbitrary orders of precision by working to higher orders in $\sigma$ (see Eq.~\eqref{neuman}).

The concept of feasibility has been associated with the extinction probability. It was summarised by Stone 1988 \cite{stone1988some} that a higher feasibility probability is linked to the reduction in the probability of extinction following structural disturbances, which are changes in interaction strengths caused by environmental change. Our results imply that increasing predator-prey interactions reduces the chance of extinction following structural disturbances.

We have used the assumption of May 1972 that all species are self-regulating. This is representative of natural ecosystems since \xl{they} require 50 percent of species to self-regulate to allow for stability \cite{barabas2017self}. However, the assumption that $r_i=1$ $\forall i\in[1,n]$ may not be biologically realistic, as natural ecosystems contain consumer species which do not grow in isolation. This is an interesting area for future investigation, however it was suggested by Song et al \cite{song2018will} that this assumption gives the parameter region where feasible systems are likely to be present.

Having generalised the distribution of $\bm{x}^{*}$ to account for arbitrary $\rho$, we have opened up the possibility for extending the results of Gibbs et al. \cite{gibbs2018effect} to analytically predict the boundary of the eigenvalue distribution of the GLV Jacobian of such systems. This would enable us to calculate the stability of such GLV models. One potential method to perform this calculation is by applying the cavity method as detailed in \cite{gibbs2018effect}. It may also be possible to calculate the expected value of $-\text{min}_{i\in\{1,n\}}x^{*}_{i}$ by applying order statistics as detailed in \cite{pettersson2020predicting}, and thus the expected resilience of a GLV model with a given value of $\rho$, although this is only applicable to systems where $|\rho|$ is small. \xl{We note, also, that the analytical approaches central to this study lead to predictions of normal distributions of steady-state species abundances. Empirical evidence is typically scale-dependent and points to a range of more complex possible species-abundance distributions \cite{antao2021shape} and the development of scale-dependent theory to bridge this gap with models may be a fruitful line of further enquiry.}

Overall, our analyses, combined with \cite{allesina2015stability,clenet2022equilibrium, pettersson2020predicting} \gc{show} that increasing the proportion of predator-prey interactions not only increases feasibility, but also the resilience of feasible GLV models. This provides greater support to Allesina and Tang's \cite{allesina2015stability} conclusion that predator-prey interactions are stabilising whilst competitive/mutualistic interactions are destabilising.

\section*{Acknowledgments}
We thank the Complexity and Stability reading group at the University of York for useful discussions.

\bibliographystyle{unsrt}
\bibliography{main.bib}

\end{document}


\title{Supplemental Material: Feasibility and Stability in Large Lotka-Volterra systems with interaction structure}

\author{Xiaoyuan Liu$^1$, George W.A. Constable, and Jonathan W. Pitchford}
\affiliation{Department of Mathematics, University of York}

\maketitle

\vfill
\setcounter{equation}{0}
\setcounter{figure}{0}
\setcounter{table}{0}
\setcounter{page}{1}
\makeatletter
\renewcommand{\theequation}{S\arabic{equation}}
\renewcommand{\thefigure}{S\arabic{figure}}

\section{Analytically Approximating $Var(x^{*}_{i})$ to Order $\sigma^6$}\label{sigma44}

\subsection{Coefficient of $\sigma^4$ in $Var(x^{*}_{i})$}\label{sigma44First}

We first approximate the coefficient of $\sigma^4$ in $Var(x^{*}_{i})$. To do this we specify the Taylor expansion of $x^{*}_{i}$ in $\sigma$ in index notation. In matrix form, the Taylor expansion of $\bm{x}^{*}$ to order $\sigma^4$ is \begin{equation}
    \bm{x}^{*}=(\mathbf{I}+\sigma\mathcal{E}+\sigma^2\mathcal{E}^2+\sigma^3\mathcal{E}^3+\sigma^4\mathcal{E}^4+ O(\sigma^5))\mathbf{r}
\label{jndfvjnfdnvjfdjkjvkdfkjv}\end{equation}
where \begin{equation}
    \mathcal{E}=\begin{pmatrix}0&a_{12}&...&a_{1n}\\&\ddots&\\&&0&\\a_{n1}&&&0\end{pmatrix}\hspace{3mm}\text{,}\hspace{5mm}\mathbf{r}=\begin{bmatrix}1\\1\\\vdots\\1\end{bmatrix}
\end{equation}
In index notation, \eqref{jndfvjnfdnvjfdjkjvkdfkjv} can be expressed as \begin{equation}
 x^{*}_{i}=1+\sigma\mathcal{E}_{ij}+\sigma^2(\mathcal{E}^2)_{ij}+\sigma^3(\mathcal{E}^3)_{ij}+\sigma^4(\mathcal{E}^4)_{ij}+O(\sigma^5)
\label{ddcdcdcdcdcdcdcd}\end{equation}
since $r_i=1$ for all $i\in[1,n]$. All terms in \eqref{ddcdcdcdcdcdcdcd} represent terms to be summed over. The subscript $i$ in \eqref{ddcdcdcdcdcdcdcd} is the free index while all other indicies are dummy indicies. Please note that terms such as $\mathcal{E}_{ij}$ denote vectors and not matrices, since $i$ is the free index. $Var(x^{*}_{i})$ is defined by the equation \begin{equation}
    Var(x^{*}_{i})=E({x^{*}_{i}}^2)-E(x^{*}_{i})^2
\label{varxmmt}\end{equation} so we need to we seek the second moment of $x^{*}_{i}$, which can be found using \eqref{ddcdcdcdcdcdcdcd}. The expression for ${x^{*}_{i}}^2$ is deduced by squaring \eqref{ddcdcdcdcdcdcdcd}. Since the expectation of all terms of odd powers of $\sigma$ is 0, we can safely ignore them, which gives \begin{equation}
    {x^{*}_{i}}^2=1+\sigma^2\big(2(\mathcal{E}^2)_{ij}+\mathcal{E}_{ij}\mathcal{E}_{ik}\big)+\sigma^4\big(2(\mathcal{E}^4)_{ij}+2(\mathcal{E}^3)_{ij}(\mathcal{E})_{ik}+(\mathcal{E}^2)_{ij}(\mathcal{E}^2)_{ik}\big)+O(\sigma^6)
\label{2ndmomen}\end{equation}
    which we can apply to calculate the second moment of $x^{*}_{i}$. We see from \eqref{2ndmomen} that we need to determine $E((\mathcal{E}^4)_{ij})$, $E((\mathcal{E}^3)_{ij}(\mathcal{E})_{ik})$ and $E((\mathcal{E}^2)_{ij}(\mathcal{E}^2)_{ik})$. The expression for $(\mathcal{E}^4)_{im}$ can be expressed as a sum of terms involving products of interaction coefficients \begin{equation}
    (\mathcal{E}^4)_{im}=\sum_{k=1}^{n}\sum_{j\neq i, j\neq k}^{n}\sum_{l\neq k,l\neq m}^{n}a_{ij}a_{jk}a_{kl}a_{lm}
\label{896}\end{equation}
and the expression for $(\mathcal{E}^4)_{ij}$ is \begin{equation}
    (\mathcal{E}^4)_{ij}=(\mathcal{E}^4)_{ii}+(\mathcal{E}^4)_{ij}\mathbf{1}_{\{j\neq i\}}
\label{141}\end{equation}
we first determine the expectation of $(\mathcal{E}^4)_{ii}r_i$. We see from \eqref{896} that the expression for $(\mathcal{E}^4)_{ii}$ is  \begin{equation}
    (\mathcal{E}^4)_{ii}=\sum_{k=1}^{n}\sum_{j\neq i, j\neq k}^{n}\sum_{l\neq k}^{n}a_{ij}a_{jk}a_{kl}a_{li}
\label{049}\end{equation} 
It is also possible for \eqref{049} to have terms where $i=k$, $j=l$ or both, which give rise to terms of \eqref{049} with nonzero expectation. To represent the case where $i=k$ but $j\neq l$, \eqref{049} is multiplied by $\delta_{ik}$ and to represent the case where $j=l$ but $i\neq k$, \eqref{049} is multiplied by $\delta_{jl}$. To represent the case where both $i=k$ and $j=l$, we multiply \eqref{049} by $\delta_{ik}\delta_{jl}$. The expectation of \eqref{049} is \begin{eqnarray}
E((\mathcal{E}^4)_{ii})&=&E\bigg(\sum_{k=1}^{n}\sum_{j\neq i, j\neq k}^{n}\sum_{l\neq k}^{n}a_{ij}a_{jk}a_{kl}a_{li}(\delta_{ik}+\delta_{jl}+\delta_{ik}\delta_{jl})\bigg)\\
&=&\sum_{j\neq i, j\neq l}^{n}\sum_{l\neq i}^{n}E(a_{ij}a_{ji}a_{il}a_{li})+\sum_{j\neq i}^{n}\sum_{k\neq j, k\neq i}^{n} E(a_{ij}a_{jk}a_{kj}a_{ji})+\sum_{j\neq i}^{n} E(a_{ij}^2a_{ji}^2)\nonumber\\&=& (n-1)(1+2(n-1)\rho^2)\label{414}\end{eqnarray}
It is also proven in Section \ref{0ex} below that the second summation term of \eqref{141} $(\mathcal{E}^4)_{ij}\mathbf{1}_{\{j\neq i\}}$ has an expectation of 0, which implies that \begin{eqnarray}E((\mathcal{E}^4)_{ij})&=&E((\mathcal{E}^4)_{ii})\label{362}\\&=&2(n-1)(n-2)\rho^2+(n-1)(1+2\rho^2)\nonumber\\&=&(n-1)(1+2(n-1)\rho^2)\nonumber\end{eqnarray}
When generalised to account for $C$, this expression becomes \begin{eqnarray}E((\mathcal{E}^4)_{ij})=(n-1)(C+\rho^2(2C+2C^2(n-2)))\label{E4C}\end{eqnarray} This generalisation is done by counting the number of pairwise uncorrelated terms in the expectation e.g in $E(a_{ij}a_{ji}a_{il}a_{li})$, the variables $(a_{ij},a_{ji})$ and $(a_{il},a_{li})$ are uncorrelated, so a factor of $C^2$ would be present in $E(a_{ij}a_{ji}a_{il}a_{li})$. It is crucial to note that \eqref{E4C} is also the coefficient of $\sigma^4$ in the expression for $E(x^{*}_{i})$. The expression for $(\mathcal{E}^3)_{ij}\mathcal{E}_{ik}$ can be expressed as a sum of terms involving products of interaction coefficients \begin{equation}
    (\mathcal{E}^3)_{ij}\mathcal{E}_{ik}=\sum_{k=1}^{n}\sum_{j\neq i, j\neq k}^{n}\sum_{l\neq k,l\neq m}^{n}a_{ij}a_{ik}a_{kl}a_{lm}
\label{9ddd6}\end{equation}
The terms that give rise to nonzero expectation of \eqref{9ddd6} include those where $m=k$ and $i=l$, $l=i$ and $m=j$ and those where $m=k$ only.
\begin{eqnarray}
E((\mathcal{E}^3)_{ij}\mathcal{E}_{ik})&=&E\bigg(\sum_{k=1}^{n}\sum_{j\neq i, j\neq k}^{n}\sum_{l\neq k}^{n}a_{ij}a_{ik}a_{kl}a_{lm}(\delta_{il}\delta_{mk}+\delta_{il}\delta_{mj}+\delta_{mk})\bigg)\\
&=&\sum_{j\neq i, j\neq l}^{n}\sum_{l\neq i}^{n}E(a_{ij}^2a_{ik}a_{ki})+\sum_{j\neq i}^{n}\sum_{k\neq j, k\neq i}^{n} E(a_{ik}^3a_{ki})+\sum_{j\neq i}^{n} E(a_{ik}^2a_{kl}a_{lk})\nonumber\\&=& 3(n-1)\rho+2(n-1)(n-2)\rho\label{hngnhngnhgjghjghj}\end{eqnarray} 

The expression for $(\mathcal{E}^2)_{ij}(\mathcal{E}^2)_{ik}$ can be expressed as a sum of terms involving products of interaction coefficients \begin{equation}
    (\mathcal{E}^2)_{ij}(\mathcal{E}^2)_{ik}=\sum_{k=1}^{n}\sum_{j\neq i, j\neq k}^{n}\sum_{l\neq k,l\neq m}^{n}a_{ij}a_{jk}a_{il}a_{lm}
\label{dnjkjkjjnkfvdfv}\end{equation}
The terms that give rise to nonzero expectation of \eqref{dnjkjkjjnkfvdfv} include those where $k=i$ and $l=j$ and $m=i$, $k=m$ and $j=l$ and those where $k=i$ and $m=i$.
\begin{eqnarray}
E((\mathcal{E}^2)_{ij}(\mathcal{E}^2)_{ik})&=&E\bigg(\sum_{k=1}^{n}\sum_{j\neq i, j\neq k}^{n}\sum_{l\neq k}^{n}a_{ij}a_{jk}a_{il}a_{lm}(\delta_{ki}\delta_{lj}\delta_{mi}+\delta_{km}\delta_{jl}+\delta_{ki}\delta_{mi})\bigg)\\
&=&\sum_{j\neq i, j\neq l}^{n}\sum_{l\neq i}^{n}E(a_{ij}^2a_{ji}^2)+\sum_{j\neq i}^{n}\sum_{k\neq j, k\neq i}^{n} E(a_{ij}^2a_{jk}^2)+\sum_{j\neq i}^{n} E(a_{ij}a_{ji}a_{il}a_{li})\nonumber\\&=& (n-1)\big((n-1)+n\rho^2\big)\label{fggbgb}\end{eqnarray} 
We have also ensured that no terms in all the summations above are double counted. Substituting all the results derived here into \eqref{2ndmomen}, we see that the coefficient of $Var(x^{*}_{i})$ at order $\sigma^4$ is $(n-1)^2+\rho(n-1)(4n-2)+\rho^2(n-1)$, and combined with the coefficient of $Var(x^{*}_{i})$ at order $\sigma^2$ we get \begin{equation}
    Var(x^{*}_{i})=(n-1)\sigma^2+\sigma^4\bigg((n-1)^2+\rho(n-1)(4n-2)+\rho^2(n-1)\bigg)+O(\sigma^6)
\end{equation}

We can generalise the expression for $Var(x^{*}_{i})$ to account for the connectance $C$. Since $a_{ij}$ and $a_{ji}$ are zero with probability $(1-C)$ and are sampled from a bivariate normal distribution with probability $C$ and $Corr(a_{ij},a_{ji})=\rho$, we have that \begin{eqnarray}
    &&Var(x^{*}_{i})=(n-1)C\sigma^2+\sigma^4\bigg([C(n-1)+C^2(n-1)(n-2)]+\rho[4C^2(n-1)(n-2)+6C(n-1)]+\nonumber\\&&\rho^2[2C(n-1)+C^2(n-1)(n-2)-(n-1)^2C^2]\bigg)+O(\sigma^6)\label{vxrhoneq0}\end{eqnarray}
The analytics in \eqref{vxrhoneq0} show that for the case where $C<1$, sampling $a_{ij}$ and $a_{ji}$ from a bivariate distribution with probability $C$ gives a different $Var(x^{*}_{i})$ to when $A$ is sampled randomly with connectance $C$, as was done by May \cite{may1972will}.

\subsection{Proof that $\sum^{n}_{j\neq i}(\mathcal{E}^4)_{ij}r_j$ has Zero Expectation}\label{0ex}

In the particular scenario where $i\neq m$ in \eqref{896}, the possible scenarios are $m=j$, $m=k$ and $m=l$. In the case where $m=j$, \eqref{896} becomes
 \begin{equation}
    (\mathcal{E}^4)_{ij}=\sum_{k=1}^{n}\sum_{j\neq i, j\neq k}^{n}\sum_{l\neq k,l\neq j}^{n}a_{ij}a_{jk}a_{kl}a_{lj}
\label{fvdvdfv}\end{equation}
and it is possible that $i=k$, $j=l$ or $i=l$. If $j=l$, \eqref{fvdvdfv} is 0 since $\mathcal{E}_{ii}=0$ for all $i\in[1,n]$. We therefore consider the cases $i=l$ and $i=k$ in the case where $m=j$. The expression for the expectation of $(\mathcal{E}^4)_{ij}$ is

\begin{eqnarray}
E((\mathcal{E}^4)_{ij})&=&E\bigg(\sum_{k=1}^{n}\sum_{j\neq i, j\neq k}^{n}\sum_{l\neq k}^{n}a_{ij}a_{jk}a_{kl}a_{lj}(\delta_{ik}+\delta_{il}+\delta_{ik}\delta_{il})\bigg)\\
&=&\sum_{j\neq i, j\neq l}\sum_{l\neq i}E(a_{ij}a_{ji}a_{il}a_{lj})+\sum_{j\neq i}\sum_{k\neq j, k\neq i} E(a_{ij}a_{jk}a_{ki}a_{ij})+0\nonumber\\&=& 0\end{eqnarray}
Now we consider the case where $m=k$. In this case, it is possible that $j=l$ and $i=l$. It is not possible for $i=k$ since we are considering the scenario in \eqref{896} where $i\neq m$. The expression for the expectation of $(\mathcal{E}^4)_{ik}$ is
\begin{eqnarray}
E((\mathcal{E}^4)_{ik})&=&E\bigg(\sum_{k=1}^{n}\sum_{j\neq i, j\neq k}^{n}\sum_{l\neq k}^{n}a_{ij}a_{jk}a_{kl}a_{lk}(\delta_{il}+\delta_{jl}+\delta_{jl}\delta_{il})\bigg)\\
&=&\sum_{j\neq i, j\neq l}\sum_{l\neq i}E(a_{ij}a_{jk}a_{ki}a_{ik})+\sum_{j\neq i}\sum_{k\neq j, k\neq i} E(a_{ij}a_{jk}a_{kj}a_{jk})+0\nonumber\\&=& 0\end{eqnarray}

Finally, in the case where $m=l$, we have that $E((\mathcal{E}^4)_{il})=0$ since $(\mathcal{E}^4)_{il}$ involves a product of $\mathcal{E}_{lm}$ which equals 0 if $l=m$. We have therefore proven that $\sum^{n}_{j\neq i}(\mathcal{E}^4)_{ij}r_j$ has an expectation of 0.

\subsection{Coefficient of $\sigma^4$ in $Cov(x^{*}_{i},x^{*}_{j})$}
Using the same technique as entailed in Section \ref{sigma44First}, here we provide the calculation of the $\sigma^4$ coefficient of $Cov(x^{*}_{i},x^{*}_{j})$. Firstly, $Cov(x^{*}_{i},x^{*}_{j})$ is defined by the equation \begin{equation}
Cov(x^{*}_{i},x^{*}_{j})=E(x^{*}_{i}x^{*}_{j})-E(x^{*}_{i})E(x^{*}_{j}).  
\label{covXIXJexpress}\end{equation}
Due to symmetry in the expression for $\bm{x}^{*}$, $E(x^{*}_{i})$ is identical to $E(x^{*}_{j})$ and so \eqref{covXIXJexpress} simplifies to \begin{equation}
Cov(x^{*}_{i},x^{*}_{j})=E(x^{*}_{i}x^{*}_{j})-E(x^{*}_{i})^2.  
\label{covXIXJexpressSimp}\end{equation}
Here, the only extra number we need to calculate is $E(x^{*}_{i}x^{*}_{j})$. Again in index notation, the expression for $x^{*}_{i}$ is given by 
\begin{equation}   x^{*}_{i}=1+\sigma\mathcal{E}_{ik}+\sigma^2(\mathcal{E}^2)_{ik}+\sigma^3(\mathcal{E}^3)_{ik}+\sigma^4(\mathcal{E}^4)_{ik}+O(\sigma^5)
\end{equation}
and $x^{*}_{j}$ is given by \begin{equation}
    x^{*}_{j}=1+\sigma\mathcal{E}_{jl}+\sigma^2(\mathcal{E}^2)_{jl}+\sigma^3(\mathcal{E}^3)_{jl}+\sigma^4(\mathcal{E}^4)_{jl}+O(\sigma^5)
\end{equation}

\begin{equation}
    x^{*}_{i}x^{*}_{j}=1+\sigma^2\big(2(\mathcal{E}^2)_{ik}+\mathcal{E}_{ik}\mathcal{E}_{jl}\big)+\sigma^4\big(2(\mathcal{E}^4)_{ik}+2(\mathcal{E}^3)_{ik}\mathcal{E}_{jl}+(\mathcal{E}^2)_{ik}(\mathcal{E}^2)_{jl}\big)
\label{xixjexpression}\end{equation}
We can calculate $E(x^{*}_{i}x^{*}_{j})$ directly from a\eqref{xixjexpression}. From \eqref{xixjexpression}, we see that the extra terms we need to determine are $E[(\mathcal{E}^3)_{ik}\mathcal{E}_{jl}]$ and $E[(\mathcal{E}^2)_{ik}(\mathcal{E}^2)_{jl}]$. $E[(\mathcal{E}^4)_{ik}]$ is already determined from our calculation of the $\sigma^4$ coefficient of $Var(x^{*}_{i})$ in Section \ref{sigma44First}. Firstly, the expression for $(\mathcal{E}^3)_{ik}\mathcal{E}_{jl}$ can be expressed as \begin{equation}
    (\mathcal{E}^3)_{ik}\mathcal{E}_{jl}=\sum_{k=1,k\neq o}^{n}\sum_{m\neq i, m\neq o}^{n}\sum_{l\neq k,l\neq m}^{n}\sum^{n}_{o\neq m,o\neq k}a_{jl}a_{im}a_{mo}a_{ok}
\label{E3IJEIKdcsdc}\end{equation}
The terms that give rise to nonzero expectation of \eqref{E3IJEIKdcsdc} include those where $j=o$ and $k=l$, $o=l$ and $k=j$, $o=i$ and $m=k$, and simply $m=k$. The expectation of \eqref{E3IJEIKdcsdc} is
\begin{eqnarray}
    E[(\mathcal{E}^3)_{ik}\mathcal{E}_{jl}]&=&E\bigg(\sum_{k=1,k\neq o}^{n}\sum_{m\neq i, m\neq o}^{n}\sum_{l\neq k,l\neq m}^{n}\sum^{n}_{o\neq m,o\neq k}a_{jl}a_{im}a_{mo}a_{ok}\big(\delta_{jo}\delta_{kl}+\delta_{lo}\delta_{kj}+\delta_{oi}\delta_{mk}+\delta_{mk}\big)\bigg)\label{E3IKEJL}\\&=&\sum^{n}_{l\neq j}\sum^{n}_{m\neq i,m\neq j}E(a_{jl}a_{im}a_{mj}a_{jl})+\sum^{n}_{l\neq j,l\neq m}\sum^{n}_{m\neq i,m\neq l}E(a_{jl}a_{im}a_{ml}a_{lj})+\sum^{n}_{k\neq i}\sum^{n}_{l\neq j}E(a_{jl}a_{ki}a_{ik}a_{ki})\nonumber\\&+&\sum^{n}_{l\neq j}\sum^{n}_{k\neq i,k\neq o}\sum^{n}_{o\neq k}E(a_{jl}a_{ik}a_{ko}a_{ok})
    \nonumber\end{eqnarray}
We now calculate the expectations of each of the individual summation terms in the second and third rows of \eqref{E3IKEJL}. \begin{equation}
    \sum^{n}_{l\neq j}\sum^{n}_{m\neq i,m\neq j}E(a_{jl}a_{im}a_{mj}a_{jl})=0
\label{exp000000}\end{equation}

\begin{eqnarray}
    \sum^{n}_{l\neq j}\sum^{n}_{m\neq i,m\neq j}E(a_{jl}a_{im}a_{ml}a_{lj})&=&\sum^{n}_{m\neq i} E(a_{ji}a_{im}a_{mi}a_{ij})+E(a_{ij}^2a_{ji}^2)\label{originalnodc}\\&=&(1+2\rho^2)+(n-2)\rho^2
\nonumber\end{eqnarray}

\begin{eqnarray}
    \sum^{n}_{l\neq j}\sum^{n}_{m\neq i,m\neq j}E(a_{jl}a_{ik}a_{ki}a_{ik})&=&E(a_{ij}^2a_{ji}^2)=(1+2\rho^2)
\label{1+2hho2_dc}\end{eqnarray}

\begin{eqnarray}
    \sum^{n}_{l\neq j}\sum^{n}_{k\neq i,l\neq n}\sum^{n}_{o\neq i}E(a_{jl}a_{ik}a_{ko}a_{ok})=\sum^{n}_{o\neq i} E(a_{ij}a_{ji}a_{io}a_{oi})=(n-2)\rho^2
\label{lastsum}\end{eqnarray}
Noting that \eqref{1+2hho2_dc} is already counted into \eqref{originalnodc}, we neglect \eqref{1+2hho2_dc} to prevent double counting. Summing \eqref{exp000000}, \eqref{originalnodc} and \eqref{lastsum} we see that \begin{equation}
  E[(\mathcal{E}^3)_{ik}\mathcal{E}_{jl}]=1+2(n-1)\rho^2  
\label{E3Efianl}\end{equation}
The expression for $(\mathcal{E}^2)_{ik}(\mathcal{E}^2)_{jl}$ is given by \begin{equation}
    (\mathcal{E}^2)_{ik}(\mathcal{E}^2)_{jl}=\sum_{k=1,k\neq m}^{n}\sum_{m\neq i, m\neq k}^{n}\sum_{l\neq o}^{n}\sum^{n}_{o\neq j,o\neq l}a_{im}a_{mk}a_{jo}a_{ol}
\label{E2IJEIKdcsdc}\end{equation}
The terms that give rise to nonzero expectation of \eqref{E2IJEIKdcsdc} include those where $m=j$ and $i=k$, $l=j$, and $m=j$ and $i=o$. The expectation of \eqref{E2IJEIKdcsdc} is 
\begin{eqnarray}
    E[(\mathcal{E}^2)_{ik}(\mathcal{E}^2)_{jl}]&=&E\bigg(\sum_{k=1,k\neq m}^{n}\sum_{m\neq i, m\neq k}^{n}\sum_{l\neq o}^{n}\sum^{n}_{o\neq j,o\neq l}a_{im}a_{mk}a_{jo}a_{ol}(\delta_{oi}\delta_{lm}+\delta_{mo}\delta_{li}+\delta_{lj})\bigg)\label{E2IKEJL}\\&=&\sum^{n}_{m\neq i,m\neq k}\sum^{n}_{k\neq m}E(a_{im}a_{mk}a_{ji}a_{im})+\sum^{n}_{m\neq i,m\neq j}\sum^{n}_{k\neq m}E(a_{im}a_{mk}a_{jm}a_{mi})+\sum^{n}_{m\neq i, m\neq k}\sum^{n}_{k\neq m}\sum^{n}_{o\neq j}E(a_{im}a_{mk}a_{jo}a_{oj})
\nonumber\end{eqnarray}

We now calculate the expectations of each individual summation term in the second line of \eqref{E2IKEJL}.

\begin{eqnarray}
    \sum^{n}_{m\neq i,m\neq k}\sum^{n}_{k\neq m}E(a_{im}a_{mk}a_{ji}a_{im})&=&E(a_{ij}^2a_{ji}^2)=(1+2\rho^2)\label{E2E21}\\
    \sum^{n}_{m\neq i,m\neq k}\sum^{n}_{k\neq m}E(a_{im}a_{mk}a_{jm}a_{mi})&=&\sum^{n}_{m\neq i,m\neq j}E(a_{im}a_{mj}a_{jm}a_{mi})=(n-2)\rho^2\label{E2E22}\\
    \sum^{n}_{m\neq i,m\neq k}\sum^{n}_{k\neq m}\sum^{n}_{o\neq j}E(a_{im}a_{mk}a_{jo}a_{oj})&=&\sum^{n}_{o\neq j}\sum^{n}_{m\neq i}E(a_{im}a_{mi}a_{jo}a_{oj})=((n-1)^2-1)\rho^2
\label{E2E23}\end{eqnarray}
summing \eqref{E2E21}, \eqref{E2E22} and \eqref{E2E23} we see that \begin{equation}
    E[(\mathcal{E}^2)_{ik}(\mathcal{E}^2)_{jl}]=1+n(n-1)\rho^2
\label{E2E2fiklal;}\end{equation}
Substituting \eqref{E2E2fiklal;} and \eqref{E3Efianl} into the expression or the expectation of \eqref{xixjexpression}, we can determine the $\sigma^4$ coefficient of $Cov(x^{*}_{i},x^{*}_{j})$ which is given by $3+5(n-1)\rho^2$. The expression for $Cov(x^{*}_{i},x^{*}_{j})$ up to and including order $\sigma^4$ is \begin{equation}
    Cov(x^{*}_{i},x^{*}_{j})=\rho\sigma^2+(3+5(n-1)\rho^2)\sigma^4
\end{equation} and for the case where $C<1$, it can be generalised to \begin{equation}
    Cov(x^{*}_{i},x^{*}_{j})=\rho C\sigma^2+(3+(6+C(5n-11))\rho^2)\sigma^4
\label{CC4}\end{equation}

\subsection{Coefficient of $\sigma^6$ in $Var(x^{*}_{i})$}\label{approx EXVXCXSIG6}

To obtain the coefficient of $\sigma^6$ in $Var(x^{*}_{i})$, we apply the Taylor expansion of $x^{*}_{i}$ up to order $\sigma^6$, given by
\begin{equation}
    \bm{x}^{*}=(\mathbf{I}+\sigma\mathcal{E}+\sigma^2\mathcal{E}^2+\sigma^3\mathcal{E}^3+\sigma^4\mathcal{E}^4+\sigma^5\mathcal{E}^5+\sigma^6\mathcal{E}^6+O(\sigma^7))\mathbf{r}
\label{neumann6th}\end{equation} which in index notation is \begin{equation}
    x^{*}_{i}=1+\sigma\mathcal{E}_{ij}+\sigma^2(\mathcal{E}^2)_{ij}+\sigma^3(\mathcal{E}^3)_{ij}+\sigma^4(\mathcal{E}^4)_{ij}+\sigma^5(\mathcal{E}^5)_{ij}+\sigma^6(\mathcal{E}^6)_{ij}+O(\sigma^7)
\label{E6a}\end{equation} Again, we need to apply \eqref{varxmmt}, which requires us to calculate the second moment of $x^{*}_{i}$ as well as $E(x^{*}_{i})^2$. To calculate $E(x^{*}_{i})$, we need to calculate $E((\mathcal{E}^6)_{ij})$. To calculate $E({x^{*}_{i}}^2)$, we need to calculate the expectation of the square of \eqref{E6a} with terms of order higher than $\sigma^6$ truncated. At order $\sigma^6$, it is convenient to adopt a different notation to that detailed in Section \ref{sigma44}. Here, we re-express the matrix $\mathcal{E}$ in the form \begin{equation}
   \mathcal{E}=\sum^{n-1}_{i=1}\sum^{n}_{j=2, j>i}\Phi_{ij}
\label{E6}\end{equation}
where $\Phi_{ij}$ represents matrices such that their $(i,j)$-th component is $a_{ij}$ (i.e. $(\Phi_{ij})_{ij}=a_{ij}$ and $(\Phi_{ij})_{ji}=a_{ji}$ for $j>i$) and all other entries are zero. The coefficients of ${x^{*}_{i}}^2$ at order $\sigma^6$ are given in \eqref{sigma6termsv} below. \begin{equation}
    2(\mathcal{E}^6)_{ij}+2(\mathcal{E}^5)_{ij}\mathcal{E}_{ik}+2(\mathcal{E}^4)_{ij}(\mathcal{E}^2)_{ik}+(\mathcal{E}^3)_{ij}(\mathcal{E}^3)_{ik}
\label{sigma6termsv}\end{equation}
and the coefficients of $E(x^{*}_{i})^2$ at order $\sigma^6$ are \begin{equation}
    2E[(\mathcal{E}^6)_{ij}]+2E[(\mathcal{E}^4)_{ii}]E[(\mathcal{E}^2)_{ii}]
\label{sigma6remaining}\end{equation}

and the coefficient of $Var(x^{*}_{i})$ at order $\sigma^6$ referred to as $v_6$ in the main text is given by

\begin{equation}
2E[(\mathcal{E}^5)_{ij}\mathcal{E}_{ik}]+2E[(\mathcal{E}^4)_{ij}(\mathcal{E}^2)_{ik}]+ E[(\mathcal{E}^3)_{ij}(\mathcal{E}^3)_{ik}]- 2E[(\mathcal{E}^4)_{ii}]E[(\mathcal{E}^2)_{ii}]  
\label{sigma6v6}\end{equation}
obtained by subtracting \eqref{sigma6remaining} from \eqref{sigma6termsv}. 
To calculate the coefficient of $\sigma^6$ in $Var(x^{*}_{i})$, it is necessary to calculate the expectation of $(\mathcal{E}^5)_{ij}(\mathcal{E})_{ik}$, $(\mathcal{E}^4)_{ij}(\mathcal{E}^2)_{ik}$ and $(\mathcal{E}^3)_{ij}(\mathcal{E}^3)_{ik}$ in \eqref{sigma6v6}. To calculate the expectations of these terms, we consider the expression for the sixth power of \eqref{E6} in terms of the $\Phi_{ij}$ terms. In Sections \ref{aaa} to \ref{ddddd}, we describe the method of calculating the expectations of each of the terms in \eqref{sigma6v6}. We note that to obtain an approximation of $E(x^{*}_{i})$ to order $\sigma^6$, it is necessary to calculate $E[(\mathcal{E}^6)_{ij}]$. The principles used to calculate this are identical to those used to calculate all the terms in \eqref{sigma6v6}, and we begin this subsection by explaining the procedures for calculating $E[(\mathcal{E}^6)_{ij}]$.

\subsubsection{Calculating $E((\mathcal{E}^6)_{ij})$}\label{aaa}

Only terms in $(\mathcal{E}^6)_{ij}$ involving products of certain $\Phi_{ij}$ terms have nonzero expectation, such as $\Phi_{12}^2\Phi_{13}^2\Phi_{23}^2$ which is the product of the square of three $\Phi_{ij}$ terms that involve mutually uncorrelated variables. As another example, terms of the form $\Phi_{ij}^4\Phi_{ik}^2$ for $j\neq k$ also has nonzero expectation. Another term in $(\mathcal{E}^6)_{ij}$ with nonzero expectation is $\Phi_{ij}^6$. To count the number of terms of each form, we consider in Table \ref{tab:my_label} below the set of all forms in which the terms having nonzero expectation can take. Let $A$, $B$ and $C$ denote the $\Phi_{ij}$ matrices with variables that are all mutually uncorrelated when $\rho\neq 0$ e.g. $A=\Phi_{12}$, $B=\Phi_{13}$ and $C=\Phi_{23}$, we have that
\begin{table}[H]
    \centering
    \begin{tabular}{|c|c|c|
} \hline       $A^2B^2C^2$  & $A^4B^2$ & $A^6$ \\
\hline       $\Phi_{ij}^2\Phi_{ik}^2\Phi_{jk}^2$  &  $\Phi_{ij}^4\Phi_{ik}^2$ &$\Phi_{ij}^6$\\ \hline$\Phi_{ij}^2\Phi_{ik}^2\Phi_{il}^2$ &$\Phi_{ij}^4\Phi_{jk}^2$&\\ \hline $\Phi_{ij}^2\Phi_{ik}^2\Phi_{kl}^2$ &$\Phi_{ij}^4\Phi_{kl}^2$&\\ \hline $\Phi_{ij}^2\Phi_{jk}^2\Phi_{kl}^2$&&\\ \hline $\Phi_{ij}^2\Phi_{ik}^2\Phi_{jl}^2$&&\\ \hline $\Phi_{ij}^2\Phi_{jk}^2\Phi_{jl}^2$ &&\\ \hline $\Phi_{ij}^2\Phi_{jk}^2\Phi_{il}^2$&&\\ \hline
    \end{tabular}
    \caption{Set of all forms of different terms of $\mathcal{E}^6$ which have nonzero expectations. We specifically do not allow any pair of indicies to be equal i.e. $i\neq j$, $j\neq k$. The first column ($A^2B^2C^2$) involves products of 3 different matrices that contain mutually uncorrelated variables, while the second column ($A^4B^2$) involves products of 2 different matrices that contain mutually uncorrelated variables. $i$ is assumed to be the free index throughout Section \ref{approx EXVXCXSIG6}, with all other indicies dummy indicies.}
    \label{tab:my_label}
\end{table}
Since matrix multiplication is not commutative, we now consider the set of all permutations of each entry of Table \ref{tab:my_label} above. First, we seek the set of all permutations of terms of each form that gives rise to a nonzero sum of first row (first row sum) (we take the first row $i=1$ W.L.O.G). $\Phi_{ij}^2\Phi_{ik}^2\Phi_{jk}^2$ contains 6 permutations that give rise to a nonzero first row sum. The expectation of the first row sum of each matrix of this form  (i.e expectation of the first row sum of $\Phi_{12}\Phi_{13}\Phi_{23}$) is $\rho^3$. Since two indicies ($j$ and $k$) are summed over, there exists $(n-1)(n-2)$ terms in the sum. There exists 6 permutations with nonzero first row sums, but three of these 6 permutations are double countings of the other three. The expectation of $\Phi_{ij}^2\Phi_{ik}^2\Phi_{jk}^2$ is thus $3(n-1)(n-2)\rho^3$.

$\Phi_{ij}^2\Phi_{ik}^2\Phi_{il}^2$ contains 2 distinct permutations that give rise to nonzero first row sum, and with 3 indicies summed over, there are $((n-1)^3-(n-1)^2-2(n-1)(n-2))$ terms in the sum. Subtraction of $(n-1)^2+2(n-1)(n-2)$ from $(n-1)^3$ is to ensure no double countings in the sum over 3 indices. Since the expectation of the first row sum of each matrix of this form is also $\rho^3$, the expectation of $\Phi_{ij}^2\Phi_{ik}^2\Phi_{il}^2$ is thus $2((n-1)^3-(n-1)^2-2(n-1)(n-2))\rho^3$. The expectation of each entry of Table \ref{tab:my_label} is deduced by following these steps.
\begin{enumerate}
  \item Checking the number of permutations of each entry of Table \ref{tab:my_label} that has a nonzero sum of (first) row. We do this for the first row if we set our dummy index $i$ to 1 W.L.O.G. Check if any pair of rows are double countings of each other.
  \item Calculating the expectation of the first row sum of each permutation that has a nonzero first row sum.
  \item Count the number of terms in the sum over its set of dummy indicies. We specifically do not allow any index to equal each other, not even the free index.
  \item Check if any terms in the sums over its dummy indicies are double counted by relabelling dummy indicies. If there are double countings, subtract the number of double counted terms from the quantity deduced in 3. For example, if summing over 2 indicies and we specifically do not allow any index to equal each other, then there are $(n-1)^2-(n-1)$ terms, since $(n-1)^2$ terms are summed over and $(n-1)$ of them are double counted.
\end{enumerate}
Finally we take the product of all quantities deduced in 1. to 3. above to give the expectation of each entry of Table \ref{tab:my_label}. Repeating this procedure for all entries of Table \ref{tab:my_label} and summing, we find that \begin{eqnarray}
    E((\mathcal{E}^6)_{ij})&=&(n-1)(n-2)+(n-1)(9\rho+6\rho^3)+6(n-1)(n-2)\rho(1+2\rho^2)\label{Ee6ij}\\&&+3(n-1)(n-2)\rho^3+5((n-1)^3-(n-1)^2-2(n-1)(n-2))\rho^3
\nonumber\end{eqnarray}
If $A$ has a connectance $C$ and $a_{ij}$ and $a_{ji}$ are sampled from a bivariate distribution, then they are both nonzero with probability $C$, and so we have \begin{eqnarray}
    E((\mathcal{E}^6)_{ij})&=&(n-1)(n-2)C^3+(n-1)(9\rho+6\rho^3)C+6(n-1)(n-2)\rho(1+2\rho^2)C^2\label{E6ij}\\&&+3(n-1)(n-2)\rho^3C^3+5((n-1)^3-(n-1)^2-2(n-1)(n-2))\rho^3C^3
\nonumber\end{eqnarray}

\subsubsection{Calculating $E((\mathcal{E}^5)_{ij}\mathcal{E}_{ik})$}

Now we seek to find $E((\mathcal{E}^5)_{ij}\mathcal{E}_{ik})$. Table \ref{tab2} shows the set of all forms in which the terms of $(\mathcal{E}^5)_{ij}\mathcal{E}_{ik}$ having nonzero expectation can take.
\begin{table}[H]
    \centering
    \begin{tabular}{|c|c|c|}
        \hline $A^2B^2C^2$&$A^4B^2$&$A^6$ \\
      \hline   $(\Phi_{ij}^2\Phi_{jk}^2\Phi_{ik})\Phi_{ik}$& $(\Phi_{ij}^4\Phi_{ik})\Phi_{ik}$&$(\Phi_{ij}^5)\Phi_{ij}$\\
      \hline $(\Phi_{ij}^2\Phi_{ik}^2\Phi_{il})\Phi_{il}$&$(\Phi_{ij}^2\Phi_{ik}^3)\Phi_{ik}$&\\
      \hline
      $(\Phi_{ij}\Phi_{jk}^2\Phi_{jl}^2)\Phi_{ij}$&$(\Phi_{ij}^3\Phi_{jk}^2)\Phi_{ij}$&\\
      \hline 
      $(\Phi_{ij}^2\Phi_{jk}^2\Phi_{il})\Phi_{il}$&$(\Phi_{ij}\Phi_{jk}^4)\Phi_{ij}$&\\
      \hline
      $(\Phi_{ij}\Phi_{ik}^2\Phi_{jl}^2)\Phi_{ij}$&&\\
      \hline
      $(\Phi_{ij}\Phi_{ik}^2\Phi_{kl}^2)\Phi_{ij}$&&
      \\
      \hline
    \end{tabular}
    \caption{Set of all forms of different terms of $(\mathcal{E}^5)_{ij}\mathcal{E}_{ik}$ which have nonzero expectations. Here, all the $\Phi_{ij}$ matrices inside the brackets represent matrices that comprise the term $(\mathcal{E}^5)_{ij}$ and the matrix outside the brackets represent the $\Phi_{ij}$ matrices that comprise $\mathcal{E}_{ik}$.}
    \label{tab2}
\end{table}
Next, we repeat step 1. detailed in the page above, but only for the terms inside the brackets. This gives $(\mathcal{E}^5)_{ij}$. After that, multiply the first row sums of all matrices with nonzero first row sum by the first row sum of the $\Phi$ outside the bracket. In other words, multiplying $(\mathcal{E}^5)_{ij}$ with $\mathcal{E}_{ik}$. We next calculate the expectation of this quantity. Then we repeat step 3 and 4. We find that \begin{eqnarray}
    E((\mathcal{E}^5)_{ij}\mathcal{E}_{ik})&=&2(n-1)(n-2)(1+2\rho^2)+5(3\rho^2(n-1)(n-2))+(n-1)(n-2)\rho\\&&+5((n-1)^3-(n-1)^2-2(n-1)(n-2))\rho^2
+(n-1)(3+12\rho^2)\nonumber\end{eqnarray}
and if $A$ has connectance $C$ with $a_{ij}$ and $a_{ji}$ sampled from a bivariate distribution, then we have
\begin{eqnarray}
    E((\mathcal{E}^5)_{ij}\mathcal{E}_{ik})&=&2(n-1)(n-2)(1+2\rho^2)C^2+4(3\rho^2(n-1)(n-2))C^2+(3\rho^2(n-1)(n-2))C^3\\&&+(n-1)(n-2)\rho C^3+5((n-1)^3-(n-1)^2-2(n-1)(n-2))\rho^2C^3
+(n-1)(3+12\rho^2)C\nonumber\end{eqnarray}

\subsubsection{Calculating $E((\mathcal{E}^4)_{ij}(\mathcal{E}^2)_{ik})$}\label{1d3}

Now we seek to find $E((\mathcal{E}^4)_{ij}(\mathcal{E}^2)_{ik})$. Table \ref{tab3} shows the set of all forms in which the terms of $(\mathcal{E}^4)_{ij}(\mathcal{E}^2)_{ik}$ having nonzero expectation can take.

\begin{table}[H]
    \centering
    \begin{tabular}{|c|c|c|}
        \hline $A^2B^2C^2$&$A^4B^2$&$A^6$ \\
      \hline   $(\Phi_{ij}^2\Phi_{jk}^2)(\Phi_{ik}^2)$& $(\Phi_{ij}^2\Phi_{ik}^2)(\Phi_{ij}^2)$&$(\Phi_{ij}^4)(\Phi_{ij}^2)$\\
      \hline $(\Phi_{ij}\Phi_{ik}\Phi_{il}^2)(\Phi_{ij}\Phi_{ik})$&$(\Phi_{ij}^2\Phi_{jk}^2)(\Phi_{ij}^2)$&\\
      \hline
       $(\Phi_{ij}\Phi_{ik}\Phi_{jk}^2)(\Phi_{ij}\Phi_{ik})$&$(\Phi_{ij}^3\Phi_{ik})(\Phi_{ij}\Phi_{ik})$&\\
      \hline 
       $(\Phi_{ij}\Phi_{ik}\Phi_{jl}^2)(\Phi_{ij}\Phi_{ik})$&$(\Phi_{ij}^3\Phi_{jk})(\Phi_{ij}\Phi_{jk})$&\\
      \hline
       $(\Phi_{ij}\Phi_{ik}\Phi_{kl}^2)(\Phi_{ij}\Phi_{ik})$&$(\Phi_{ij}\Phi_{jk}^3)(\Phi_{ij}\Phi_{jk})$&\\
      \hline
       $(\Phi_{ij}\Phi_{jk}\Phi_{ik}^2)(\Phi_{ij}\Phi_{jk})$&$(\Phi_{ij}^4)(\Phi_{ik}^2)$&
      \\
      \hline
       $(\Phi_{ij}\Phi_{jk}\Phi_{il}^2)(\Phi_{ij}\Phi_{jk})$&&
       \\
       \hline
        $(\Phi_{ij}\Phi_{jk}\Phi_{jl}^2)(\Phi_{ij}\Phi_{jk})$&&
        \\
        \hline
         $(\Phi_{ij}\Phi_{jk}\Phi_{kl}^2)(\Phi_{ij}\Phi_{jk})$&&
         \\
         \hline
          $(\Phi_{ij}^2\Phi_{jk}^2)(\Phi_{il}^2)$&&
          \\
          \hline
          $(\Phi_{ij}^2\Phi_{ik}^2)(\Phi_{il}^2)$&&
          \\
          \hline
    \end{tabular}
    \caption{Set of all forms of different terms of $(\mathcal{E}^4)_{ij}(\mathcal{E}^2)_{ik}$ with nonzero expectations. Here, all the $\Phi_{ij}$ matrices inside the left bracket represent those that comprise the $(\mathcal{E}^4)_{ij}$ term and all the $\Phi_{ij}$s in the right bracket represents all those in the $(\mathcal{E}^2)_{ik}$ term.}
    \label{tab3}
\end{table}
Now, it is necessary to seek the set of all $(\mathcal{E}^4)_{ij}(\mathcal{E}^2)_{ik}$ such that both $(\mathcal{E}^4)_{ij}$s and $(\mathcal{E}^2)_{ik}$s have nonzero first row sums. In other words, all the entries in Table \ref{tab3} in which both their left and right brackets have nonzero first row sums. To do this, we repeat step 1 for both the left and right brackets of each entry of table \ref{tab3}, discarding any permutations of matrices in each bracket that have zero first row sums. Next, calculate the expectation of the product of the first row sum of the left bracket and the first row sum of the right bracket. Finally, repeat steps 3. and 4. We find that \begin{eqnarray}
    E((\mathcal{E}^4)_{ij}(\mathcal{E}^2)_{ik})&=&4(n-1)(n-2)\rho(1+2\rho^2)+8\rho(n-1)(n-2)\\&&+3\rho((n-1)^3-(n-1)^2-2(n-1)(n-2))+(n-1)(n-2)\rho^3+(n-1)(n-2)\rho^2\nonumber\\&&+2((n-1)^3-(n-1)^2-2(n-1)(n-2))\rho^3+(n-1)(9\rho+6\rho^3)\nonumber
\end{eqnarray}
and if $A$ has connectance $C$ with $a_{ij}$ and $a_{ji}$ sampled from a bivariate distribution, then we have
\begin{eqnarray}
    E((\mathcal{E}^4)_{ij}(\mathcal{E}^2)_{ik})&=&4(n-1)(n-2)\rho(1+2\rho^2)C^2+6\rho(n-1)(n-2)C^2+2\rho(n-1)(n-2)C^3\\&&+3\rho((n-1)^3-(n-1)^2-2(n-1)(n-2))C^3+(n-1)(n-2)\rho^3C^3+(n-1)(n-2)\rho^2C^3\nonumber\\&&+2((n-1)^3-(n-1)^2-2(n-1)(n-2))\rho^3C^3+(n-1)(9\rho+6\rho^3)C\nonumber
\end{eqnarray}
Now, we find $E((\mathcal{E}^3)_{ij}(\mathcal{E}^3)_{ik})$. Since $(\mathcal{E}^3)_{ij}(\mathcal{E}^3)_{ik}$ involves the product of two terms that contain a third power of $\mathcal{E}$, each one of which has a range of permutations that give rise to nonzero first row, $(\mathcal{E}^3)_{ij}(\mathcal{E}^3)_{ik}$ will have more permutations that give nonzero first row. For instance, the first $(\mathcal{E}^3)_{ij}$ can contain a product of either two or three different $\Phi_{ij}$s. We therefore follow an alternative set of steps. 
\begin{enumerate}
    \item Find the set of all permutations of the tuples $(1,2,3,1,2,3)$ and $(1,1,1,1,2,2)$. Each number in the tuple represents a distinct $\Phi_{ij}$ matrix e.g. 1 representing $\Phi_{ij}$, 2 for $\Phi_{ik}$, 3 for $\Phi_{il}$. Tuple $(1,2,3,1,2,3)$ represents the order of products of three distinct matrices $(A^2B^2C^2)$ while $(1,1,1,1,2,2)$ represents the order of products of two distinct matrices $(A^4B^2)$. The permutations of each of these tuples should be expressed as a table.
    \item Split the tables of permutations found in step 1. into two tables by separating columns 1 to 3 and 4 to 6. In each of these two tables, discard any repeated rows if any. The table containing columns 1 to 3 represents the set of all permutations of $(\mathcal{E}^3)_{ij}$ and the table containing columns 4 to 6 represents the set of all permutations of $(\mathcal{E}^3)_{ik}$.
    \item Check which rows of both tables represent permutations that give nonzero first row sum in BOTH $(\mathcal{E}^3)_{ij}$ and $(\mathcal{E}^3)_{ik}$.
    \item Repeat Steps 3 and 4 in the list above \eqref{Ee6ij}.
    
    \end{enumerate}
    
\subsubsection{Calculating $E((\mathcal{E}^3)_{ij}(\mathcal{E}^3)_{ik})$}\label{ddddd}
The entries of Table \ref{tab4} represents the set of all $\Phi_{ij}$ matrices that comprise the $(\mathcal{E}^4)_{ij}(\mathcal{E}^2)_{ik}$s which have nonzero expectations.
    
    \begin{table}[H]
    \centering
    \begin{tabular}{|c|c|c|}
        \hline $A^2B^2C^2$&$A^4B^2$&$A^6$ \\
      \hline  $\Phi_{ij}\Phi_{ik}\Phi_{il}$ & $\Phi_{ij}\Phi_{ik}$ &$\Phi_{ij}$\\
      \hline $\Phi_{ij}\Phi_{ik}\Phi_{jk}$ &$\Phi_{ij}\Phi_{jk}$&\\
      \hline
       $\Phi_{ij}\Phi_{ik}\Phi_{jl}$&&\\
      \hline 
       $\Phi_{ij}\Phi_{jk}\Phi_{jl}$&&\\
      \hline
       $\Phi_{ij}\Phi_{jk}\Phi_{kl}$&&\\
      \hline
       $\Phi_{ij}\Phi_{ik}\Phi_{kl}$&&
      \\
          \hline
    \end{tabular}
    \caption{Each entry of the table contains the set of  $\Phi_{ij}$ matrices that make up the $(\mathcal{E}^4)_{ij}(\mathcal{E}^2)_{ik}$ with nonzero expectations. The column $A^2B^2C^2$ are the set of products of three different $\Phi_{ij}$ matrices that can give nonzero first row sums depending on the permutation. $A^4B^2$ are the set of products of two different $\Phi_{ij}$ matrices that can give nonzero first row sums, and $A^6$ is the product of a one such matrix e.g. $\Phi_{ij}^6$.}
    \label{tab4}
\end{table}
Using the method outline immediately above, we find that
\begin{eqnarray}
    E((\mathcal{E}^3)_{ij}(\mathcal{E}^3)_{ik})&=&4\rho^2((n-1)^3-(n-1)^2-2(n-1)(n-2))+2(n-1)(n-2)\rho^2\\&&+(n-1)(n-2)\rho^3+(n-1)(n-2)+((n-1)^3-(n-1)^2-2(n-1)(n-2))\nonumber\\&&+4(n-1)(n-2)3\rho^2+2(n-1)(n-2)(1+2\rho^2)+(n-1)(3+12\rho^2)
\nonumber\end{eqnarray}
and if $A$ has connectance $C$ with $a_{ij}$ and $a_{ji}$ sampled from a bivariate distribution, then we have
\begin{eqnarray}
    E((\mathcal{E}^3)_{ij}(\mathcal{E}^3)_{ik})&=&4\rho^2((n-1)^3-(n-1)^2-2(n-1)(n-2))C^3+2(n-1)(n-2)\rho^2C^3\label{cccccccccccccccc}\\&&+(n-1)(n-2)\rho^3C^3+(n-1)(n-2)C^3+((n-1)^3-(n-1)^2-2(n-1)(n-2))C^3\nonumber\\&&+4(n-1)(n-2)3\rho^2C^2+2(n-1)(n-2)(1+2\rho^2)C^2+(n-1)(3+12\rho^2)C
\nonumber\end{eqnarray}
We can use \eqref{Ee6ij} to \eqref{cccccccccccccccc} to calculate \eqref{sigma6v6}. Putting this together, we see that the order $\sigma^6$ coefficient of $Var(x^{*}_{i})$ is equal to \begin{eqnarray}
   && C(n-1) (9 - 12 C - 2 C^2 (n-2) + 6 C n + C^2 (n-2) n)\label{sigma6v6explicit}\\&&+ 
 C (n-1) \big(18 - 38 C - 12 C^2 (n-2) + 18 C n + 
    6 C^2 (n-2) n\big)\rho \nonumber\\&&+ 
 C (n-1) (36 - 96 C - 32 C^2 (n-2) + 48 C n + 14 C^2 (n-2) n) \rho^2 \nonumber\\&&+ 
 C (n-1) (12 - 28C - 5C^2 (n-2) + 12 C n) \rho^3
\nonumber\end{eqnarray}

\subsubsection{Number of Extra terms required to calculate $Cov(x^{*}_{i},x^{*}_{j})$ to order $\sigma^6$}\label{extratermscov}

We note that the fact that the expression for $Cov(x^{*}_{i},x^{*}_{j})$ involves an extra (free) index $j$ makes the algebra required for its calculation lengthier. We have the additional term $\Phi_{ij}$ where both $i$ and $j$ are the free indices. For clarity, in this section we will denote the second free index as $i'$ instead of $j$. The number of indices inside each bracket of the tables in Section \ref{approx EXVXCXSIG6} is at most the number of unique $\Phi$ plus 1 (i.e $(\Phi_{ij}\Phi_{ik}\Phi_{il}^2)$ in Section \ref{1d3} has 3 unique $\Phi$s $\Phi_{ij}$, $\Phi_{ik}$ and $\Phi_{il}$, and 4 indices $i$,$j$,$k$ and $l$). We formalise this as \begin{equation}
    N^{max}_{ind}=N_{uniq\Phi}+1
\end{equation} It is possible for the index $i'$ to occur in the bracket a various number of times. The maximum number of times $i'$ can occur in the left bracket is \begin{equation}
    max_{(\#i')}=N^{max}_{ind}-1
\end{equation}
since the left bracket has to include the free index $i$. Just for terms of the form $(A^2B^2)C^2$ (see Table in Section \ref{1d3}), the number of terms needed is equal to 1 plus the number of terms of the form $(A^2B^2)C^2$ already present in $Var(x^{*}_{i})$ $(N^{Var(x^{*}_{i})}_{(A^2B^2)C^2})$ all times $max_{(\# i')}$, in other words \begin{equation}
    (N^{Var(x^{*}_{i})}_{(A^2B^2)C^2}+1)max_{(\#j)}
\end{equation} When we have two free indices $i$ and $i'$, $N^{max}_{ind}=3$ and $max_{(\# i')}=2$. In the expression for $Var(x^{*}_{i})$, there are 3 terms of the form $(A^2B^2)C^2$ so $N^{Var(x^{*}_{i})}_{(A^2B^2)C^2}=3$. Since $max_{(\# i')}=2$, we have $(N^{Var(x^{*}_{i})}_{(A^2B^2)C^2}+1)max_{(\#i')}=8$, and so we require 8 terms of the form $(A^2B^2)C^2$. Since it is also possible for the index $i'$ to not occur inside the bracket $(A^2B^2)$, we have an extra $(N^{Var(x^{*}_{i})}_{(A^2B^2)C^2}+1)$ terms which brings the total number of $(A^2B^2)C^2$ type terms to 12. The terms of the form $(A^2B^2)C^2$ are $(\Phi_{ij}^2\Phi_{jk}^2)(\Phi_{i'j}^2)$, $(\Phi_{ij}^2\Phi_{ik}^2)(\Phi_{i'j}^2)$, $(\Phi_{ij}^2\Phi_{ii'}^2)(\Phi_{i'j}^2)$, $(\Phi_{ii'}^2\Phi_{ij}^2)(\Phi_{i'k}^2)$, $(\Phi_{ii'}^2\Phi_{i'j}^2)(\Phi_{i'k}^2)$, $(\Phi_{ij}^2\Phi_{i'j}^2)(\Phi_{i'k}^2)$, $(\Phi_{ij}^2\Phi_{ik}^2)(\Phi_{i'l}^2)$, $(\Phi_{ij}^2\Phi_{jk}^2)(\Phi_{i'l}^2)$, $(\Phi_{ij}^2\Phi_{ik}^2)(\Phi_{i'l}^2)$, $(\Phi_{ij}^2\Phi_{jk}^2)(\Phi_{i'i}^2)$, $(\Phi_{ij}^2\Phi_{ik}^2)(\Phi_{i'i}^2)$, $(\Phi_{ij}^2\Phi_{i'j}^2)(\Phi_{i'i}^2)$.

Terms of the form $(AB^2C^2)(A)$ contain 3 variables in the left bracket, and so it can have 4 indices in that bracket. This means that $N^{max}_{ind}=4$ and $max_{(\#j)}=3$. In the expression for $Var(x^{*}_{i})$, there are 6 terms of the form $(AB^2C^2)A$, so $N^{max}_{ind}=6$, and we require $(6+1)\times 3$ = 21 terms.

\subsection{Accuracy of Analytical Prediction}
The analytical prediction of $P_{feas}(\gamma,\rho)$ remains accurate up to $|\rho|=0.5$ for $n=100$ and $|\rho|=0.25$ for $n=25$. 

\renewcommand{\thefigure}{S\arabic{figure}}
\setcounter{figure}{0}

\begin{figure}[H]
    \centering
    \includegraphics[width=100mm]{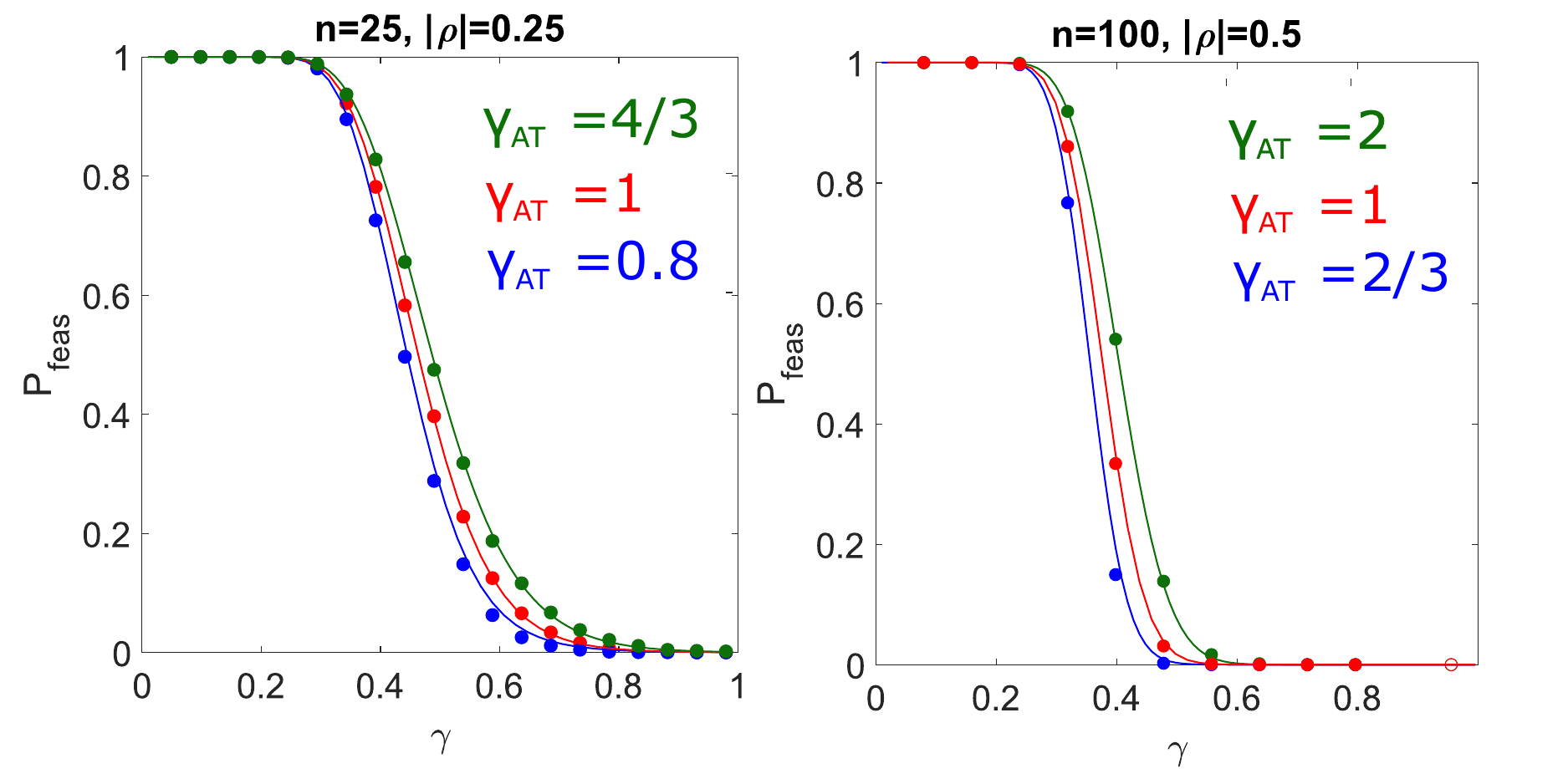}
    \caption{Dots are numerical simulations of $P_{feas}(\gamma,\rho)$. Solid curves are analytical predictions of $P_{feas}(\gamma,\rho)$ using $E(x^{*}_{i})$ approximated up to and including order $\sigma^3$ and $Var(x^{*}_{i})$ up to and including order $\sigma^6$. The system has $C=1$. Blue $\rho<0$, red $\rho=0$ and green $\rho>0$. $\gamma_{AT}$ is the complexity above which linear stability is lost in the Allesina and Tang 2015 model. For any value of $\rho$, feasibility is lost at smaller complexities than linear stability in large systems.}
    \label{fi}
\end{figure}

\begin{figure}[H]
    \centering
    \includegraphics[width=110mm]{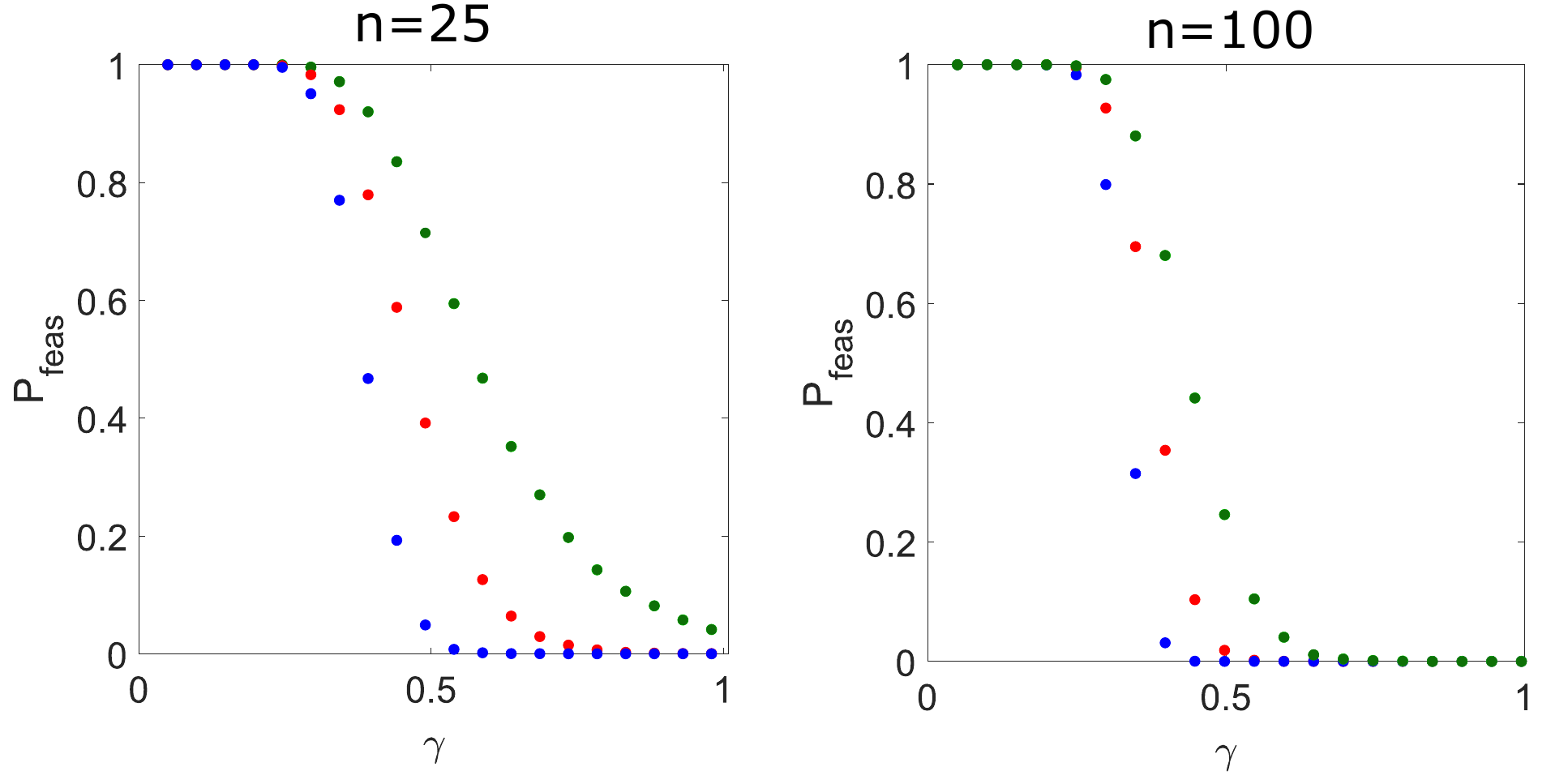}
    \caption{Numerical simulations of $P_{feas}(\gamma,\rho)$ for the case where $|\rho|=1$. Analytical predictions break down for this magnitude of $\rho$, so only numerical simulations are included. The system has $C=1$. }
    \label{fig:kjlklbel}
\end{figure}

We see from Figure \ref{fi} that feasibility is lost at a smaller complexity compared to linear stability. For systems with $\rho<0$, feasibility is lost at a much smaller complexity compared to linear stability than systems with $\rho>0$, implying that increasing the proportion of predator-prey interactions will have a more modest effect on stability than predicted by Allesina and Tang. Figure \ref{fig:kjlklbel} plots numerical simulations of $P_{feas}(\gamma,\rho)$ for the case where $\rho=\pm 1$. For the case where $\rho=-1$, $P_{feas}$ still decreases to 0 above a sufficiently large complexity even though the system is linearly stable for all complexities. 

\section{Parameter Regions Where Approximation of $Var(x^{*}_{i})$ Beyond Order $\sigma^6$  is Required}\label{breaks}

\begin{figure}[H]
    \centering
    \includegraphics[width=100mm]{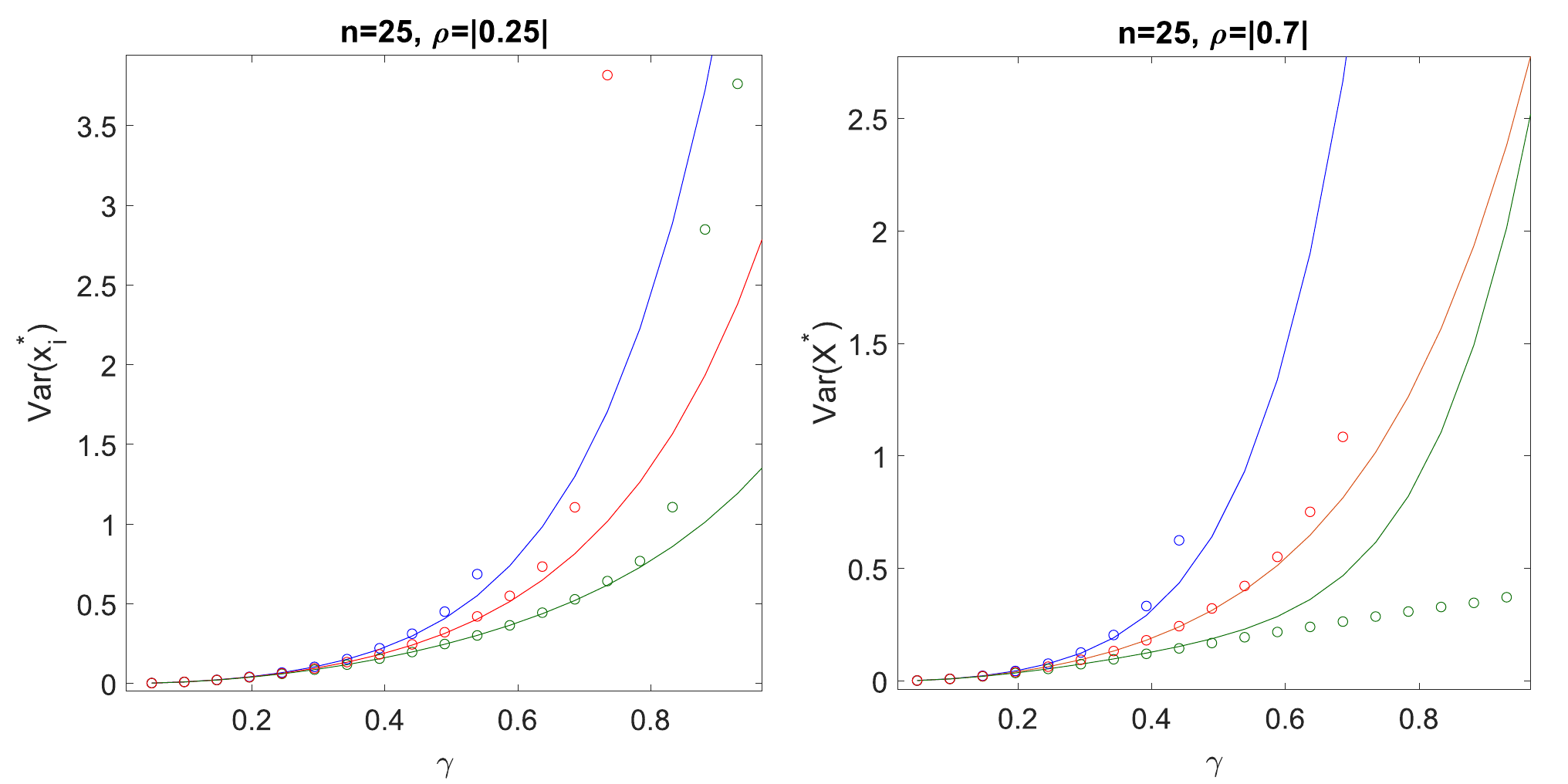}
    \caption{$Var(x^{*}_{i})$ as a function of $\gamma$ for systems with $n=25$. Panel (a) is for system with $|\rho|=0.25$ and panel (b) for $|\rho|=0.7$. It is evident from panel (b) that for large negative $\rho$ such as $\rho=-0.7$, the analytical prediction of $Var(x^{*}_{i})$ breaks down at around $\gamma=0.5$, which is a smaller value than for the case $\rho=-0.25$ where the analytical prediction breaks down at a around $\gamma=0.8$. For $\rho=0$ and $\rho>0$, outliers in $x^{*}_{i}$ begin to emerge above a sufficiently large $\gamma$, causing numerical data of $Var(x^{*}_{i})$ to become noisy.}
    \label{fvfvfdbhjvedthnf }
\end{figure}

\begin{figure}[H]
    \centering
    \includegraphics[width=100mm]{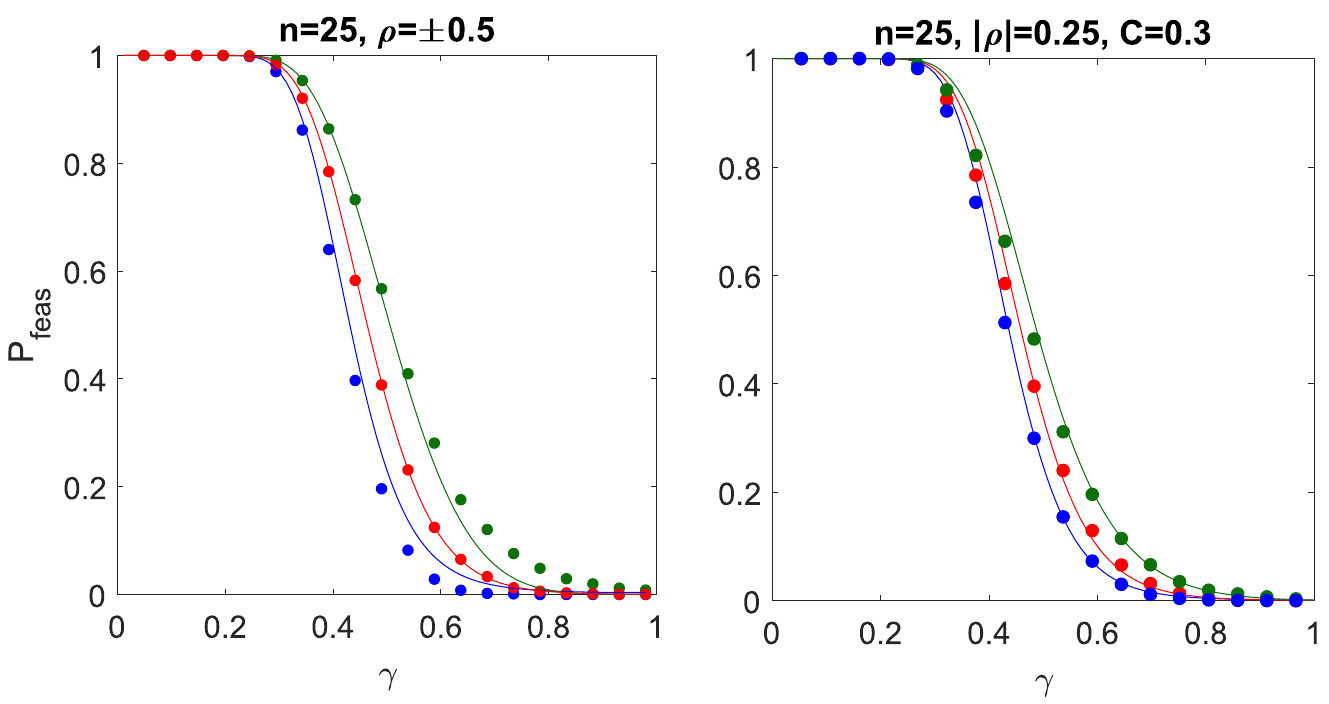}
    \caption{Left: The analytical prediction of $P_{feas}(\gamma,\rho)$ breaks down when $\rho$ is sufficiently large and negative given a sufficiently small $n$. Here $C=1$. Right: The analytical prediction of $P_{feas}(\gamma,\rho)$ begins to break down when $C$ is sufficiently small.}
    \label{whercjnkcl}
\end{figure}

In Figure \ref{whercjnkcl}, we show examples of plots in parameter regions where the approximation of $P_{feas}(\gamma,\rho)$ breaks down. It is apparent from Figure \ref{fvfvfdbhjvedthnf } that for large magnitudes of $\rho$, the analytical approximation of $Var(x^{*}_{i})$ to order $\sigma^6$ breaks down at a smaller value of $\sigma$. Since systems with a large negative $\rho$ also has a higher $P_{feas}$ at a given value of $\sigma$, the analytical prediction of $P_{feas}(\gamma,\rho)$ becomes inaccurate before $P_{feas}(\gamma,\rho)$ transitions to 0. These points indicate how the analytical approximation would break down given a sufficiently large negative $\rho$. We also show in Figure \ref{whercjnkcl} that for systems such as $n=25$, the analytical prediction of $P_{feas}(\gamma,\rho)$ also breaks down for small $C$.

In the right panel ($n=25$, $\rho=-0.7$), the feasibility probability at value of $\gamma$ at which the order $\sigma^6$ approximation of $Var(x^{*}_{i})$ breaks down is 0.632, whereas in the left panel ($n=25$, $\rho=-0.25$), this corresponding feasibility probability is 0.021. In Figure \ref{whercjnkcl}, we show that the analytical prediction of $P_{feas}(\gamma,\rho)$ can break down if either $\rho$ is sufficiently large or if $C$ is sufficiently small for a fixed community size $n$.
 It is worth noting that when $\rho$ becomes sufficiently large and negative, the variance-covariance matrix ceases to be positive semi-definite if $Var(x^{*}_{i})$ is approximated up to order $\sigma^6$.
 

\section{Analytical Prediction of Feasibility Probability as a Function of Complexity}

The analytical prediction of $P_{feas}$ as a function of complexity $\gamma$ is determined by integrating the joint density function of $\bm{x}^{*}$ from $x^{*}_{i}=0$ to $x^{*}_{i}=\infty$ for all $i\in[1,n]$. For systems of $n\leq25$, this is done using the \textit{mvncdf} command in MATLAB. For systems where $n>25$, \textit{mvncdf} is no longer applicable, therefore we obtain the analytical prediction of $P_{feas}$ as a function of $\gamma$ by reducing the multivariate normal integral to a single integral using the method detailed in \cite{curnow1962numerical}. The multivariate normal distribution function is given by \begin{equation}
    F_{\mathbf{X}}(\mathbf{X})=\int^{x_i}_{-\infty}\hdots\int^{x_n}_{-\infty}f_\mathbf{X}(\mathbf{X},\Sigma^{\mathbf{X}})d\mathbf{X}
\end{equation}
where $\Sigma^{\mathbf{X}}$ is the variance-covariance matrix of $\mathbf{X}$. Here $\mathbf{X}$ is a random variable such that $X_i=-x^{*}_{i}$. Define $y_i$ as a standardised normal random variable $y_i=\frac{X_i-\mu_{X_i}}{\sigma_{X_i}}$. If $Corr(x^{*}_{i},x^{*}_{j})$ can be expressed in the form $Corr(x^{*}_{i},x^{*}_{j})=b_ib_j$ where $b_i,b_j\in\mathbb{R}$, then $F_\mathbf{X}(\mathbf{X})$ can be expressed as \begin{equation}
    F_\mathbf{X}(\mathbf{X})=\int^{\infty}_{-\infty}\bigg\{\prod^{n}_{i=1}\Phi(\frac{y_i-b_iu}{(1-b_i^2)^{1/2}})\bigg\}\phi(u)du
\label{fhpercubeSI}\end{equation}
where $\phi(u)$ is the density function of a standard normal random variable $u$ and $\Phi(v)$ denotes the cumulative distribution function of a standard normal random variable $v$ \cite{curnow1962numerical}. The condition of feasibility $\bm{x}^{*}>\bm{0}$ is equivalent to the condition that $\mathbf{X}<\bm{0}$. Since $y_i=\frac{X_i-E(X_i)}{\sigma_{X_i}}$, the condition that $X_i<0$ is equivalent to the condition that $y_i<-E(X_i)/\sigma_{X_i}$ and thus $y_i<E(x^{*}_i)/\sigma_{x^{*}_i}$. In our analytical prediction of $P_{feas}$, we have that $y_i=\frac{E(x^{*}_i)}{\sqrt{Var(x^{*}_{i})}}$ and $b_i=\frac{\sqrt{Cov(x^{*}_{i},x^{*}_{j})}}{Var(x^{*}_{i})}$. In other words, $P_{feas}$ is the expression you get by plugging these expressions for $y_i$ and
$b_i$ into \eqref{fhpercubeSI}. The integral \eqref{fhpercubeSI} is also applicable for $Cov(x^{*}_{i},x^{*}_{j})<0$, however the integrand is complex \cite{curnow1962numerical}. 
As a result, we approximated $P_{feas}(\gamma,\rho)$ for the case $\rho<0$ by numerically integrating \eqref{fhpercubeSI} using the \textit{Nintegrate} command in mathematica. The lower and upper limits of this integral are set to -20 and 20 respectively. This numerical integral works well since the imaginary part is of magnitude $10^{-18}$ and the real part matches the numerical simulations of $P_{feas}(\gamma,\rho)$ closely, as can be seen in Figure \ref{fi}.

\section{Obtaining Numerical Simulations of Feasibility Probability}\label{Nnn}

In fully connected systems $C=1$, we obtained 10000 numerical solutions of $\bm{x}^{*}$ by doing 10000 successive runs of the GLV equation (Eq.~(5) in main text). \xl{In particular, the numerical solutions were obtained by generating 10000 random interaction matrices $A$ parameterised according to Eq.~(3) of the main text and calculating each of their corresponding $\bm{x}^{*}$ using Eq.~(5).} \xl{The feasibility probability corresponds to the fraction of numerical solutions of $\bm{x}^{*}$ out of the 10000 \xl{realisations} that contain all non-negative entries (i.e. the fraction of numerical solutions of $\bm{x}^{*}$ such that $x^{*}_{i}>0$ $\forall$ $i\in[1,n]$)}. \xl{For each value of $n$, $\rho$ and $C$ we considered, the step described above is repeated for a range of values of $\sigma$, where this range falls within the interval $\gamma\in[0,1]$ (where $\gamma=\sigma\sqrt{nC}$).} For predator-prey, mutualistic and competitive interactions ($\rho\neq 0$), the interaction matrix $A$ is constructed according to Eq.~(3) of the main text. For fully connected ($C=1$) systems with random interaction structure (i.e Figure (3) red in main text), $A$ is constructed according to Eq.~(2) of the main text.

When considering sparsely connected matrices ($C<1$), we exclude such matrices $A$ that contain disconnected components. In particular, we obtain 10000 realisations of $\bm{x}^{*}$ which correspond to sparse interaction matrices constructed according to Eq.~(3) and contain no disconnected components. Even in the case where $\rho=0$, we construct $A$ according to Eq.~(3) rather than Eq.~(2). This is done because for the case where $C<1$, constructing $A$ according to Eq.~(3) would mean that if $A_{ij}$ is nonzero, then $A_{ji}$ is also nonzero, which is not true if $A$ is constructed according to Eq.~(2). As a result, these two different ways of constructing $A$ would give different $P_{feas}$. However in the case where $C=1$, constructing $A$ according to Eq.~(3) with $\rho=0$ would give the same feasibility probability as when $A$ is constructed according to Eq.~(2).

\section{Analytical Prediction for Sparsely Connected Systems}\label{Accounting for Connectance}

\begin{figure}[H]
    \centering
    \includegraphics[width=60mm]{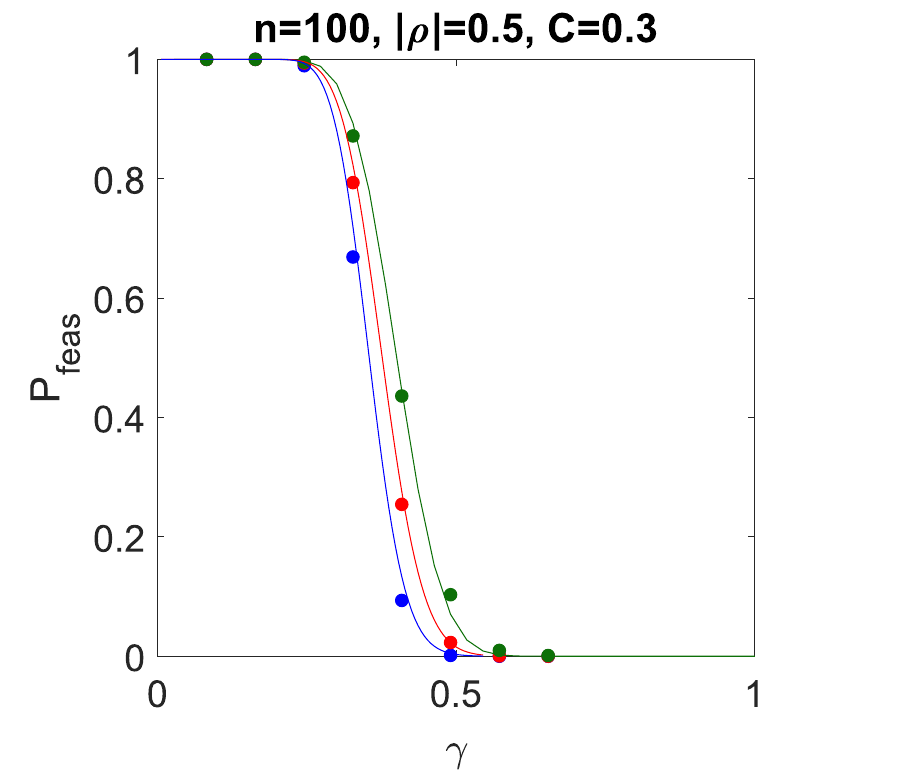}
    \caption{We show that the analytical prediction of feasibility probability as a function of complexity \cite{stone1988some} can be generalised for $C$ as well as $\rho$ (i.e. $P_{feas}$ as a funtion of $\gamma$ where $\gamma=\sigma\sqrt{nC}$). All labels and parameters for this figure are the same as that of Figure 3 right, except $C=0.3$. Our analysis of $n$ species systems concerns systems comprising $n$ species that interact as a single unit, so interaction matrices of $C<1$ that contain disconnected components are excluded from our analysis (see Supplemental Information \ref{Nnn}).}
    \label{c000000}
\end{figure}

Empirical ecological networks may be sparsely connected \cite{gardner1970connectance}, so it would be useful to generalise our feasibility calculations to account for connectance $C$. Recent work by Akjouj and Najim \cite{akjouj2021feasibility} have shown that in sparse random GLV models with block structure, $P_{\text{feas}}$ also exhibits a rapid transition to 0 above a critical interaction strength, hinting that it could be viable to generalise Stone's analytical prediction to account for $C$. In this section, we demonstrate the success of this generalisation by providing an example in Figure \ref{c000000}. In Figure \ref{c000000}, we show that for a system with $n=100$, analytical predictions for $P_{feas}(\gamma,\rho)$ remain highly accurate even in systems with connectance as low as $C=0.3$.

By comparing the analytical and numerical data in Figure \ref{c000000} with those of Figure \ref{fi}, we see that the system with $C=0.3$ shows an almost identical feasibility profile with a system of $C=1$. In other words, we get the same value of $P_{feas}$ for a given value of $\gamma_M$ in both systems. This implies that increasing $\sigma$ and decreasing $C$ to give the same complexity has negligible effect on the feasibility profile.

\section{Effect of $\rho$ on Outlier Eigenvalue}

This section shows the effect of $\rho$ on the stability of GLV models by looking at how the outlier eigenvalue of $J$ changes with $\rho$. It also shows that the abundance of the least abundant species is a good predictor of the outlier eigenvalue of $J$, and thus its stability.

\begin{figure}[H]
    \centering
    \includegraphics[width=150mm]{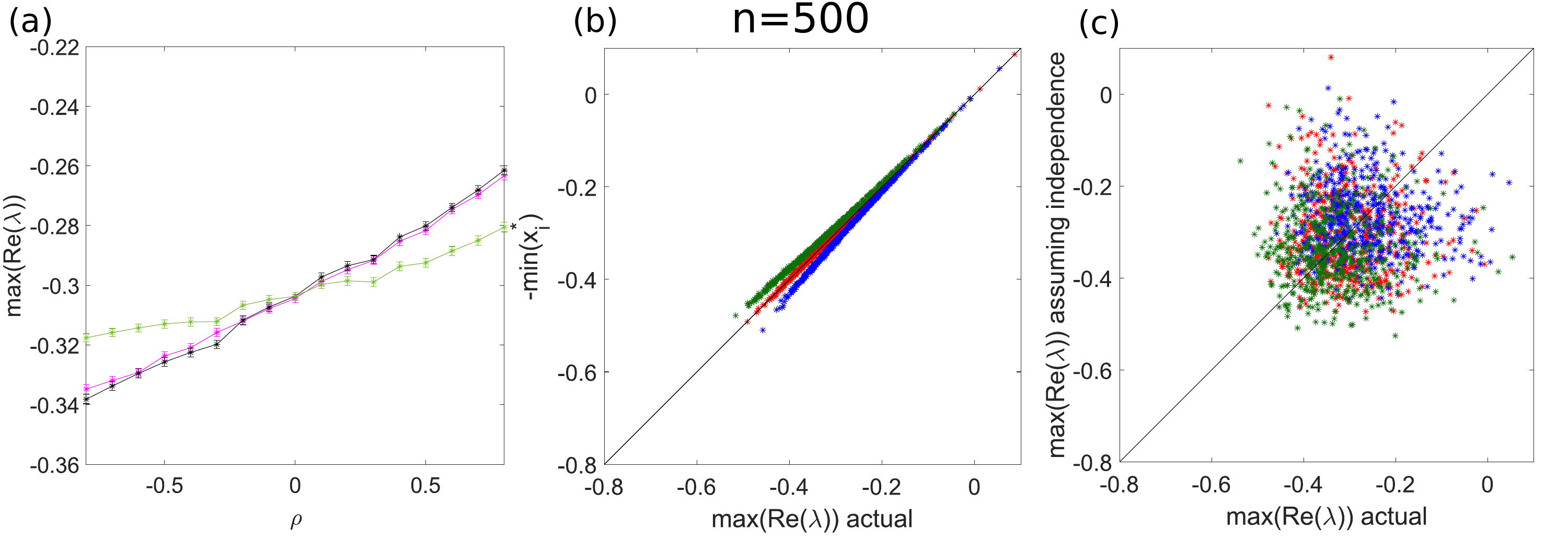}
    \caption{Panel (a) shows the effect of $\rho$ on the outlier eigenvalue of $J=\bm{x}^{*}A$, averaged over 3500 realisations. Black represents the outlier eigenvalue of the actual GLV Jacobian ($\text{max}(Re(\lambda))$ actual), \xl{where each realisation possesses a feasible $\bm{x}^{*}$}. Light green represents the outlier eigenvalue approximated by the relation $\text{max}(Re(\lambda))=-\text{min}_{i\in\{1,n\}}x^{*}_{i}$ and pink is the outlier eigenvalue of $J$ constructed by sampling $\bm{x}^{*}$ and $A$ independently. Error bars represent the standard error about the mean. Panel (b) plots $-\text{min}_{i\in\{1,n\}}x^{*}_{i}$ against $\text{max}(Re(\lambda))$ actual for 500 realisations of the GLV model, with $\gamma=0.01\sqrt{500}$. Blue $\rho=0.7$, red $\rho=0$ and green $\rho=-0.7$. Black line is the line on which $max(Re(\lambda))=-\text{min}_{i\in\{1,n\}}x^{*}_{i}$. Panel (c) plots $\text{max}(Re(\lambda))$ actual against that of $J$ constructed by sampling $\bm{x}^{*}$ independently of $A$.}
    \label{lambdamax_rho&xmin_lambdamax}
\end{figure}

Panel (a) shows that the outlier of the Jacobian constructed by sampling $x^{*}$ independently of $A$ (Grilli's assumption) correctly captures the qualitative effect of $\rho$ on stability. Although it is shown in panel (c) that constructing the Jacobian by adopting Grilli's assumption fails to accurately calculate the outlier eigenvalue of each Jacobian. Panel (b) shows that $-\text{min}_{i\in\{1,n\}}x^{*}_{i}$ is a highly accurate predictor of stability of the GLV model corresponding to a given realisation of $A$, since most green markers sit close to the diagonal line. Notice that for systems where $\rho>0$, the markers lie below the diagonal line. This implies that the stability is marginally overestimated by the relation $max(Re(\lambda))=-\text{min}_{i\in\{1,n\}}x^{*}_{i}$ and for $\rho<0$, this relation underestimates the stability slightly. Since $Corr(\text{max}(Re(\lambda)),-\text{min}_{i\in\{1,n\}}x^{*}_{i})=0.9999$, when $\rho=0$ $Corr(\text{max}(Re(\lambda)),-\text{min}_{i\in\{1,n\}}x^{*}_{i})=0.9996$ when $\rho=-0.7$ and $Corr(\text{max}(Re(\lambda)),-\text{min}_{i\in\{1,n\}}x^{*}_{i})=0.9995$ when $\rho=0.7$, $-\text{min}_{i\in\{1,n\}}x^{*}_{i}$ is an accurate predictor of stability for all regimes of $\rho$. It is of note that for large magnitudes of $\rho$, $-\text{min}_{i\in\{1,n\}}x^{*}_{i}$ becomes a poor predictor of the outlier eigenvalue of $J$ statistically, and thus a poor estimator of stability.

For smaller $n$ systems, the accuracy of $-\text{min}_{i\in\{1,n\}}x^{*}_{i}$ at predicting the outlier eigenvalue of $J$ is reduced to such an extent that it ceases to accurately predict the effect of $\rho$ on resilience.

\section{Effect of $E(x^{*}_{i})$, $Var(x^{*}_{i})$ and $Cov(x^{*}_{i},x^{*}_{j})$ on Probability of Feasibility}

For a multivariate normal distribution with a given $E(x^{*}_{i})$ and $Cov(x^{*}_{i},x^{*}_{j})$, Increasing $Var(x^{*}_{i})$ decreases feasibility probability. Increasing $E(x^{*}_{i})$ acts to increase $P_{\mathrm{feas}}$, given fixed values of $Cov(x^{*}_{i},x^{*}_{j})$ and $Var(x^{*}_{i})$ and increasing $Cov(x^{*}_{i},x^{*}_{j})$ also acts to increase $P_{\mathrm{feas}}$ given fixed values of $Var(x^{*}_{i})$ and $E(x^{*}_{i})$. In Figure \ref{EXVXCX_sensitive} below, we show that out of these three quantities, $Cov(x^{*}_{i},x^{*}_{j})$ increases most slowly with $\sigma$ (and thus $\gamma$). As a result, $Cov(x^{*}_{i},x^{*}_{j})$ plays the smallest part in governing how $P_{\text{feas}}$ changes with $\gamma$. In Figure \ref{Pfeas_gamma_sensitivity_s9}, we justify how this argument holds true by quantifying the effects of $E(x^{*}_{i})$, $Var(x^{*}_{i})$ and $Cov(x^{*}_{i},x^{*}_{j})$ on $P_{feas}$.

\begin{figure}[H]
    \centering
    \includegraphics[width=150mm]{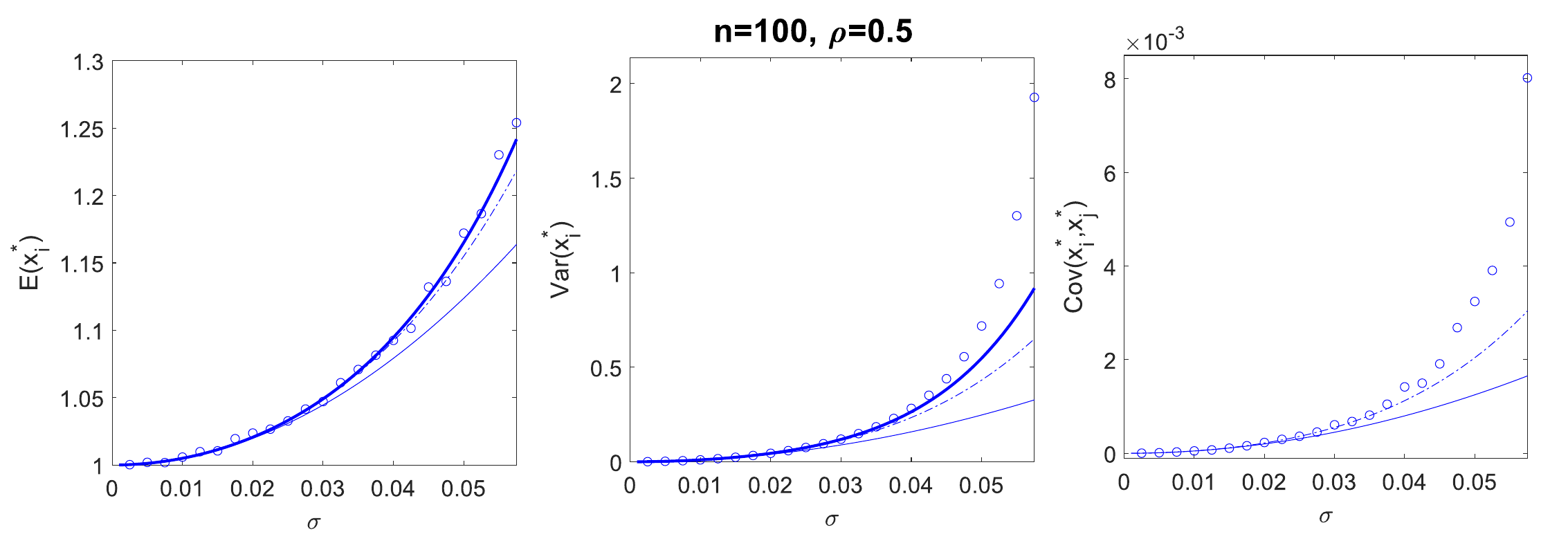}
    \caption{Analytical approximations of $E(x^{*}_{i})$, $Var(x^{*}_{i})$ and $Cov(x^{*}_{i},x^{*}_{j})$ as a function of $\sigma$ at various orders of $\sigma$. Fine solid curve order $\sigma^2$, dash dotted curve order $\sigma^4$ and bold solid curve order $\sigma^6$. Circles are numerical simulations of these quantities, obtained from 10000 numerical solutions of main text Eq.~(5), which are acquired as described in Section IV. Here, $C=1$. Since we do not have an analytical approximation of $Cov(x^{*}_{i},x^{*}_{j})$ to order $\sigma^6$, there is no bold solid curve in the right panel. For values of $\sigma$ such that $\sigma\sqrt{nC}>1/(1+\rho)$, the Neumann approximation of $\bm{x}^{*}$ Eq.~(9) breaks down given fixed $n$ and $C$. This condition is equivalent to $\sigma>0.0670$ here. Since $n$ is finite, the normality in distribution of $\bm{x}^{*}$ breaks down at some point where $\sigma\sqrt{nC}<1/(1+\rho)$ (see Section \ref{Distofxstar} for explanation). Due to this, numerical results for $Var(x^{*}_{i})$ and $Cov(x^{*}_{i},x^{*}_{j})$ no longer converges upon increase in sample size. $\sigma$ is plotted up to the largest value in which these numerical results still converge, which is 0.0575 here.}
    \label{EXVXCX_sensitive}
\end{figure}

In Figure \ref{Pfeas_gamma_sensitivity_s9} below, we show how varying each of the quantities $E(x^{*}_{i})$, $Var(x^{*}_{i})$ and $Cov(x^{*}_{i},x^{*}_{j})$ individually impacts $P_{feas}$. In each panel, we vary one of the three quantities while keeping the other two fixed. In the middle panel of Figure \ref{Pfeas_gamma_sensitivity_s9}, we see that if we vary $Var(x^{*}_{i})$ by an amount equal to the difference between its approximation at order $\sigma^2$ and $\sigma^6$, $P_{feas}$ changes by 0.5493, which is significantly large. Even if we vary $Var(x^{*}_{i})$ by the difference between its approximation at order $\sigma^4$ and $\sigma^6$, $P_{feas}$ still changes by 0.1097. If we vary  $Cov(x^{*}_{i},x^{*}_{j})$ by an amount equal to the difference between its approximation at order $\sigma^2$ and $\sigma^4$, an even smaller change in $P_{feas}$ manifests (change of 0.0019). Since we do not have the approximation of $Cov(x^{*}_{i},x^{*}_{j})$ at order $\sigma^6$, we showed what happens if we vary $Cov(x^{*}_{i},x^{*}_{j})$ by an amount equal to the difference between its approximation at order $\sigma^2$ and its actual value. In this case, $P_{feas}$ exhibits a change of 0.0037. All of the above demonstrate that going to higher orders in the approximation $Cov(x^{*}_{i},x^{*}_{j})$ confers negligible effect on $P_{feas}$.

\begin{figure}[H]
    \centering
    \includegraphics[width=150mm]{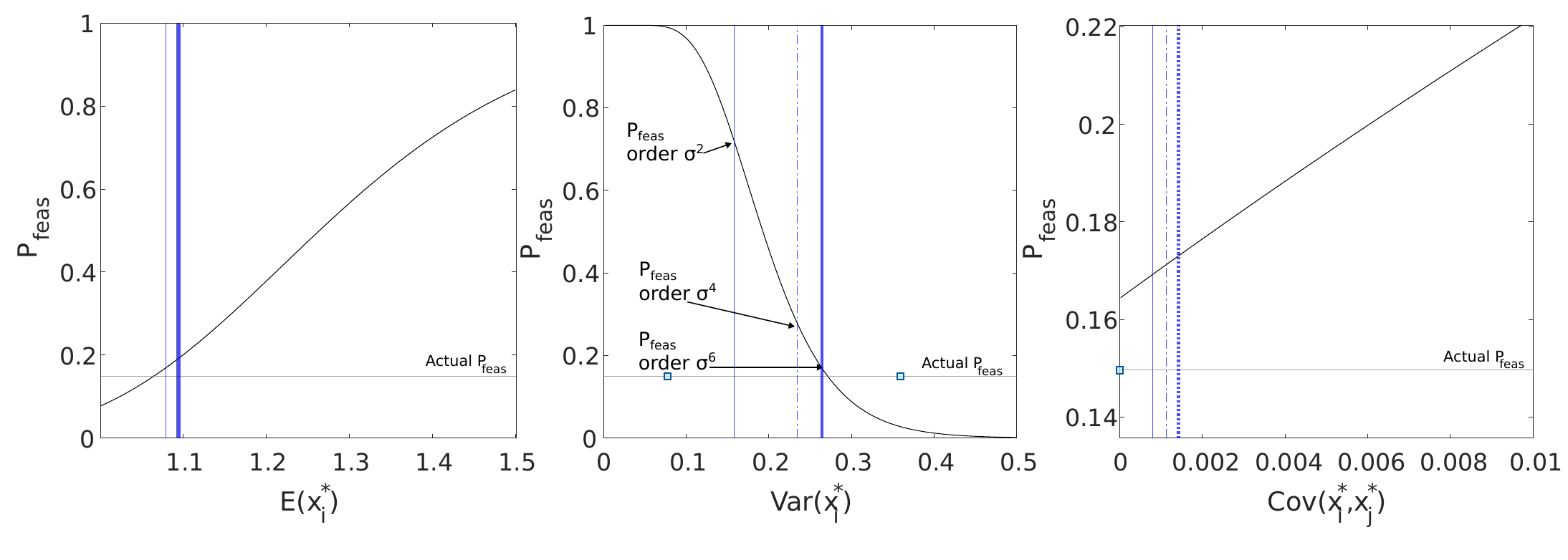}
    \caption{Panels showing how $P_{feas}$ changes as each quantity ($E(x^{*}_{i})$, $Var(x^{*}_{i})$ or $Cov(x^{*}_{i},x^{*}_{j})$) is varied, provided that the other two quantities are fixed, e.g. if $E(x^{*}_{i})$ is varied, we fix $Var(x^{*}_{i})$ and $Cov(x^{*}_{i},x^{*}_{j})$. The two fixed quantities are set to the analytically approximated value (Eq.~(12)-(14) in main text) they would take if $\sigma=0.04$, $n=100$ and $C=1$. Black curve shows how $P_{feas}$ varies with each of these quantities and vertical lines show the varying quantity analytically approximated at different orders of $\sigma$. Notations for the line textures are consistent with all other Figures in this section, except here, we have a dotted line which represents the actual numerically simulated $Cov(x^{*}_{i},x^{*}_{j})$, since we do not have the analytically approximation of $Cov(x^{*}_{i},x^{*}_{j})$ at order $\sigma^6$.}
    \label{Pfeas_gamma_sensitivity_s9}
\end{figure}
Below, we show analytically that $Cov(x^{*}_{i},x^{*}_{j})$ increases more slowly with $\sigma$ compared to $E(x^{*}_{i})$ and $Var(x^{*}_{i})$.

\subsection*{Analytical Illustration of the Magnitudes of $E(x^{*}_{i})$, $Var(x^{*}_{i})$ and $Cov(x^{*}_{i},x^{*}_{j})$}

Here, we illustrate analytically how $Cov(x^{*}_{i},x^{*}_{j})$ increases more slowly with $\sigma$ than $E(x^{*}_{i})$ and $Var(x^{*}_{i})$, and therefore plays the smalest part in governing how $P_{feas}$ changes with $\sigma$. In the expression for $Var(x^{*}_{i})$, the coefficient of $\sigma^2$ contains a term of order $n$, and the corresponding coefficient in the expression for $E(x^{*}_{i})$ contains a term of order $n\rho$. The corresponding coefficient in $Cov(x^{*}_{i},x^{*}_{j})$ contains a term of order $\rho$. At order $\sigma^4$, the coefficient for $Var(x^{*}_{i})$ includes terms of order $n^2$, $\rho n^2$ and $\rho^2n$ while the coefficient of $E(x^{*}_{i})$ includes terms of order $n$ and $\rho^2 n^2$. Finally, the coefficient of $Cov(x^{*}_{i},x^{*}_{j})$ includes terms of order 1 and $\rho^2 n$. We see that at both order $\sigma^2$ and $\sigma^4$, the coefficients of $Cov(x^{*}_{i},x^{*}_{j})$ are a factor $n$ smaller than those of $E(x^{*}_{i})$ and $Var(x^{*}_{i})$, which implies that $Cov(x^{*}_{i},x^{*}_{j})$ increases slowly with $\sigma$ given fixed values of $n$ and $C$. The small increase in $Cov(x^{*}_{i},x^{*}_{j})$ with $\sigma$ is also shown numerically in Figure \ref{Pfeas_gamma_sensitivity_s9}.

\section{Distribution of $x^{*}_{i}$}\label{Distofxstar}
It is crucial to note that the Neumann series in Eq.~(9) of main text is convergent if $||\sigma\mathcal{E}||<1$. This corresponds to the spectral radius of $A$ being less than 1. For systems of large $n$, this spectral radius is analytically approximated by $\sigma\sqrt{nC}(1+\rho)$ (see Eq.~(4) of main text), and thus the Neumann propagation of $x^{*}_{i}$ is applicable only under the condition $\sigma\sqrt{nC}<1/(1+\rho)$ (or equivalently $\gamma<1/(1+\rho)$). Stone \cite{stone1988some} argued using the Central limit theorem (CLT) that Eq.~(9) is normally distributed as $n\to\infty$. This implies that the statement that $x^{*}_i$ is normal for all cases satisfying $\sigma\sqrt{nC}<1/(1+\rho)$ is only exact in the limit as $n\to\infty$. Here we show using numerical solutions to Eq.~(5) that the distribution of $x^{*}_{i}$ is normal for small $\sigma$, and this normality breaks down for larger values of $\sigma$. The numerical solutions are obtained as described in Section \ref{Nnn}. 

\begin{figure}[H]
    \centering
    \includegraphics[width=110mm]{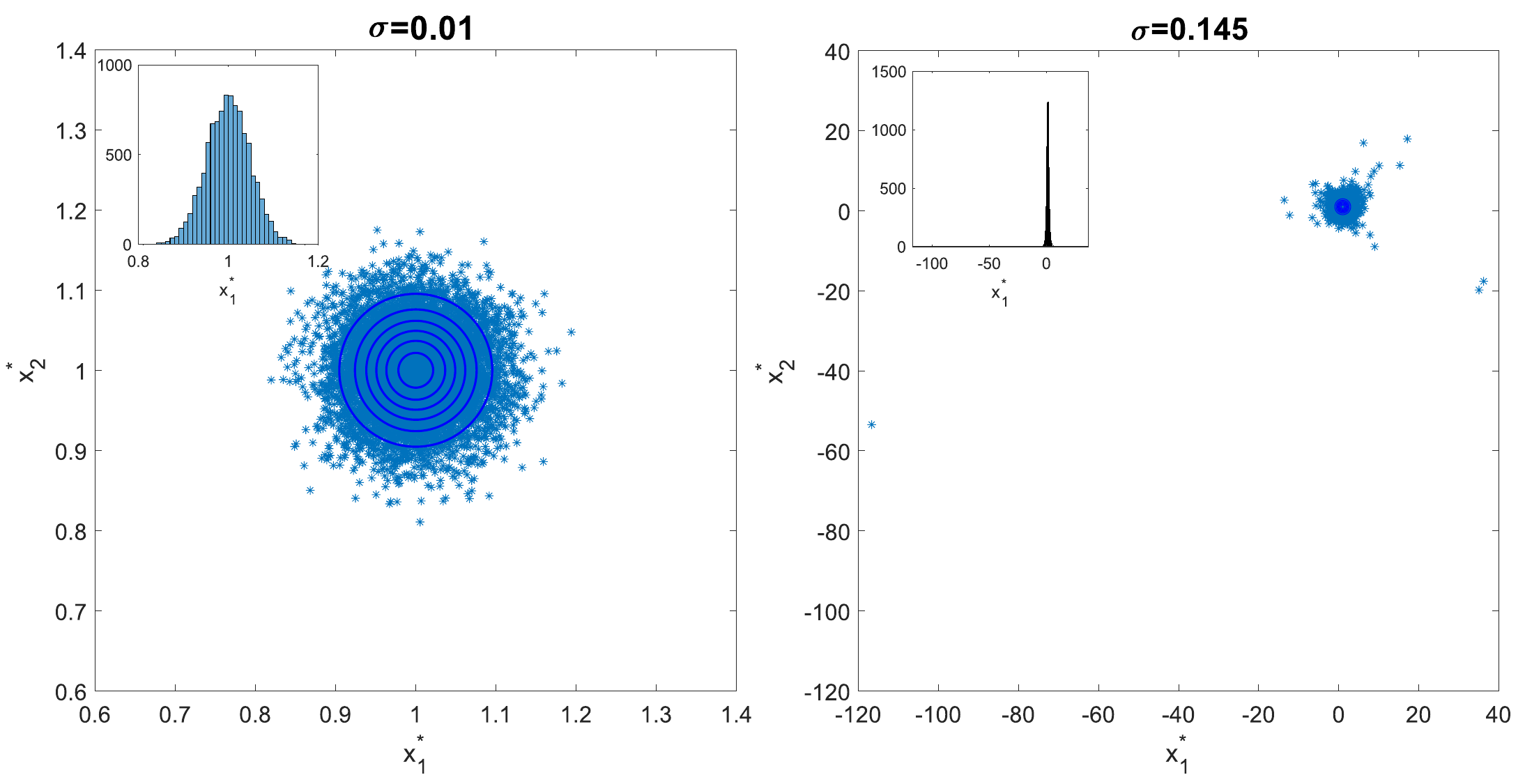}
    \caption{For an $n=25$ system, $x^{*}_{i}$ is normally distributed for small values of $\sigma$, such as $\sigma=0.01$. For this given system size $n=25$, the normality breaks down when $\sigma=0.145$, which corresponds to $\gamma=0.725$. Light blue markers represent $x^{*}_{1}$ and $x^{*}_{2}$ values of 10000 numerical solutions of $\bm{x}^{*}$, obtained as described in Section \ref{Nnn}. Panel insets show histograms for the distribution of $x^{*}_{1}$. Other parameters are $\rho=0$ and $C=1$.}    \label{xstar_n25_normality}
\end{figure}

\begin{figure}[H]
    \centering
    \includegraphics[width=110mm]{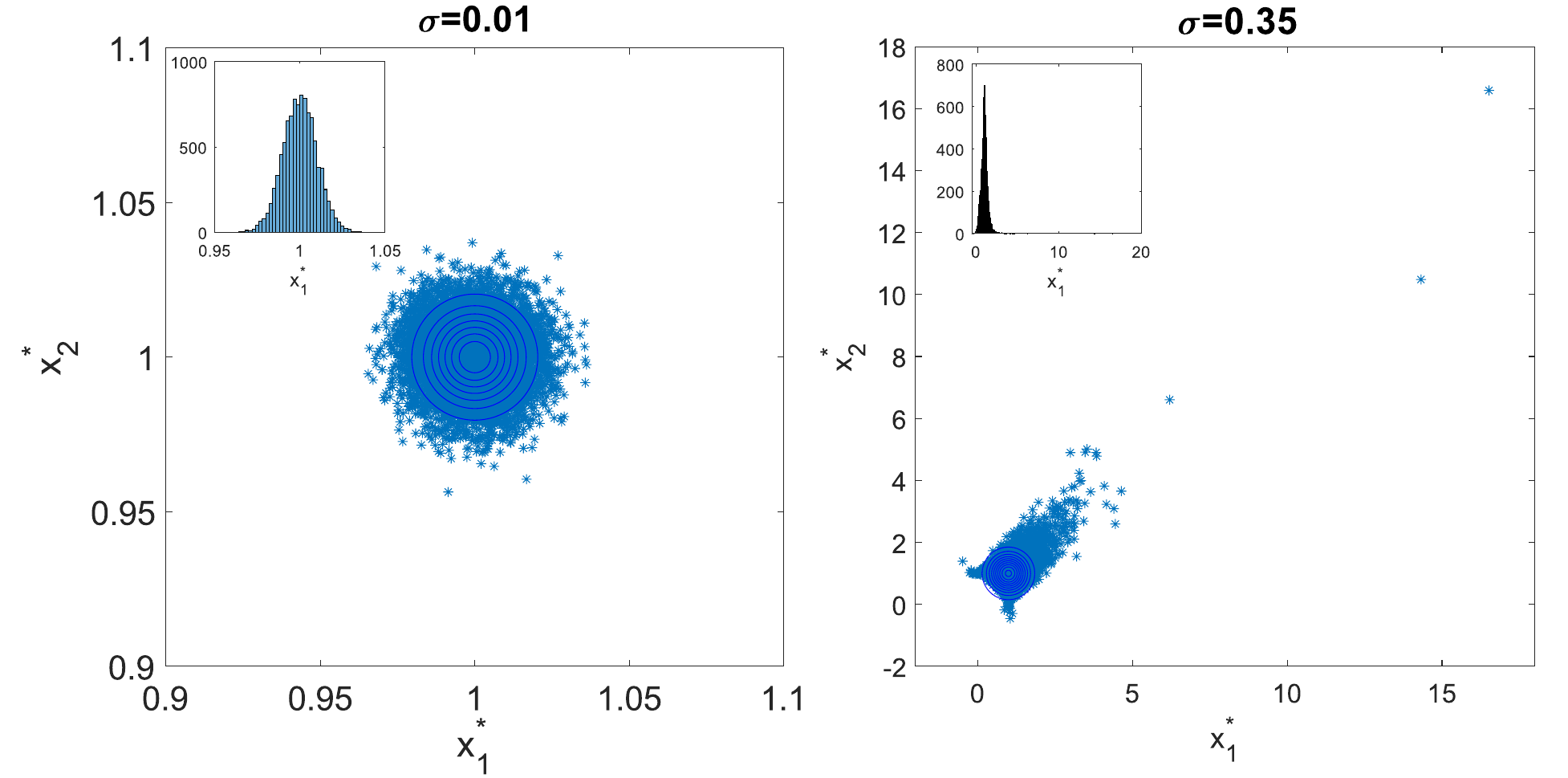}
    \caption{As in Figure \ref{xstar_n25_normality} but for an $n=2$ system. For an $n=2$ system, the normality breaks down when $\sigma=0.35$, which corresponds to $\gamma=0.495$.}
    \label{xstar_n2_normality}
\end{figure}

This agrees with the CLT argument above which infers that the larger the value of $n$, the smaller the value of $\sigma$ at which the boundary $\sigma\sqrt{nC}=1/(1+\rho)$ is reached (provided a given $C$), and thus the distribution of $x^{*}_{i}$ is more likely to be normal at or near the boundary $\gamma=1/(1+\rho)$ as we increase $\sigma$. As we increase $n$, the normality in the distribution of $x^{*}_{i}$ breaks down at a larger value of $\gamma$ such that $\gamma<1/(1+\rho)$ (see Figure \ref{xstar_n25_normality} and \ref{xstar_n2_normality}). In theory, as $n\to\infty$, the distribution of $x^{*}_{i}$ would remain normal up to $\gamma=1/(1+\rho)$.

We see from Figure \ref{xstar_n25_normality} and \ref{xstar_n2_normality} that as $n$ increases, the value of $\gamma$ at which normality in $x^{*}_{i}$ is lost increases. Similarly for the equivalent $n=100$ system as in the two Figures above, numerical results show that normality is lost at $\sigma=0.085$ which corresponds to $\gamma=0.85$. We see that the larger the value of $n$, the larger the value of $\gamma$ up to which the distribution of $x^{*}_{i}$ remains normal. In theory, as $n\to\infty$, $x^{*}_{i}$ would remain normally distributed up to $\gamma=1/(1+\rho)$, which corresponds to $\gamma=1$ for the scenario in the two Figures above, since we have $\rho=0$. Our analytical results for $P_{feas}$ in Figure (3) of the main text are highly accurate since for systems where $n\geq 25$, $P_{feas}$ already becomes approximately 0 before the point at which normality in $\bm{x}^{*}$ is lost (see Figure \ref{fi}). At $\gamma=0.725$ where normality in $\bm{x}^{*}$ breaks down for an $n=25$, $\rho=0$ system, $P_{feas}\approx 0.02$.

\bibliographystyle{unsrt}
\bibliography{supplement.bib}